\documentclass[twocolumn]{aastex631}
\usepackage{CJKutf8} 
\usepackage{amsmath}
\usepackage{comment}
\usepackage{physics}
\usepackage{verbatim}

\usepackage{longtable}

\usepackage{soul}

\submitjournal{AJ}

\begin{document}

\title{LIGHTS. Survey Overview and a Search for Low Surface Brightness Satellite Galaxies}
  
\correspondingauthor{Dennis Zaritsky}
\email{dennis.zaritsky@gmail.com}

\author[0000-0002-5177-727X]{Dennis Zaritsky}
\affiliation{Steward Observatory and Department of Astronomy, University of Arizona, 933 N. Cherry Ave., Tucson, AZ 85721, USA}

\author[0009-0001-2377-272X]{Giulia Golini}
\affiliation{Instituto de Astrof\'sica de Canarias, c/ V\'ia L\'actea s/n, 38205 La Laguna, Tenerife, Spain}
\affiliation{Departamento de Astrofísica, Universidad de La Laguna, 38206 La Laguna, Tenerife, Spain}

\author[0000-0001-7618-8212]{Richard Donnerstein}
\affiliation{Steward Observatory and Department of Astronomy, University of Arizona, 933 N. Cherry Ave., Tucson, AZ 85721, USA}

\author[0000-0001-8647-2874]{Ignacio Trujillo}
\affiliation{Instituto de Astrof\'sica de Canarias, c/ V\'ia L\'actea s/n, 38205 La Laguna, Tenerife, Spain}
\affiliation{Departamento de Astrofísica, Universidad de La Laguna, 38206 La Laguna, Tenerife, Spain}

\author[0000-0003-3449-2288]{Mohammad Akhlaghi}
\affiliation{Centro de Estudios de F\'isica del Cosmos de Arag\'on (CEFCA), Plaza San Juan, 1, E-44001, Teruel, Spain}

\author[0000-0002-1598-5995]{Nushkia Chamba}
\affiliation{NASA Ames Research Center, Moffett Field, CA 94035, USA}

\author[0000-0001-8647-2874]{Mauro D'Onofrio}
\affiliation{Department of Physics and Astronomy, University of Padua, Vicolo Osservatorio 3, 35122 Padova, Italy}

\author[0000-0002-6672-1199]{Sepideh Eskandarlou}
\affiliation{Centro de Estudios de F\'isica del Cosmos de Arag\'on (CEFCA), Plaza San Juan, 1, E-44001, Teruel, Spain}

\author[0000-0003-3449-2288]{S.Zahra Hosseini-ShahiSavandi}
\affiliation{Department of Physics and Astronomy, University of Padua, Vicolo Osservatorio 3, 35122 Padova, Italy}

\author[0000-0002-6220-7133]{Ra\'ul Infante-Sainz}
\affiliation{Centro de Estudios de F\'isica del Cosmos de Arag\'on (CEFCA), Plaza San Juan, 1, E-44001, Teruel, Spain}

\author[0000-0003-2939-8668]{Garreth Martin}
\affiliation{School of Physics and Astronomy, University of Nottingham, University Park, Nottingham NG7 2RD, UK}

\author[0000-0001-7847-0393]{Mireia Montes}
\affiliation{Instituto de Astrof\'sica de Canarias, c/ V\'ia L\'actea s/n, 38205 La Laguna, Tenerife, Spain}
\affiliation{Departamento de Astrofísica, Universidad de La Laguna, 38206 La Laguna, Tenerife, Spain}

\author[0000-0002-3849-3467]{Javier Román}
\affiliation{Kapteyn Astronomical Institute, University of Groningen, PO Box 800, 9700 AV Groningen, The Netherlands}
\affiliation{Departamento de Astrofísica, Universidad de La Laguna, 38206 La Laguna, Tenerife, Spain}

\author[0009-0001-9574-8585]{Nafise Sedighi}
\affiliation{Department of Physics, Yazd University, 8915818411, Yazd, Iran}

\author[0009-0004-5054-5946]{Zahra Sharbaf}
\affiliation{Instituto de Astrof\'sica de Canarias, c/ V\'ia L\'actea s/n, 38205 La Laguna,
Tenerife, Spain}
\affiliation{Departamento de Astrofísica, Universidad de La Laguna, 38206 La Laguna, 
Tenerife, Spain}

\begin{abstract}
We present an overview of the LIGHTS (LBT Imaging of Galactic Halos and Tidal Structures) survey, which currently includes 25 nearby galaxies that are on average $\sim$ 1 mag fainter than the Milky Way, and a catalog of 54 low central surface brightness (24 $< \mu_{0,g}$/mag arcsec$^{-2} < 28$) satellite galaxy candidates, most of which were previously uncatalogued. 
The depth of the imaging exceeds the full 10-year depth of the Rubin Observatory's Legacy Survey of Space and Time (LSST). 
We find, after applying completeness corrections, rising numbers of candidate satellites as we approach the limiting luminosity (M$_r \sim -8$ mag) and central surface brightness ($\mu_{0,g} \sim 28$ mag arcsec$^{-2}$). Over the parameter range we explore, each host galaxy (excluding those that are in overdense regions, apparently groups) has nearly 4 such candidate satellites to a projected radius of $\sim$ 100 kpc. These objects are mostly just at or beyond the reach of spectroscopy unless they are H I rich or have ongoing star formation. We identify 3, possibly 4, ultra-diffuse satellite galaxies (UDGs; effective radius $ > 1.5$ kpc). This incidence rate falls within expectations of the extrapolation of the published relationship between the number of UDGs and host halo mass. Lastly, we visually identify 12 candidate satellites that host a nuclear star cluster (NSC). The NSC occupation fraction for the sample (12/54) matches that published for satellites of early-type galaxies, suggesting that the parent's morphological type plays at most a limited role in determining the NSC occupation fraction. 

\end{abstract}


\section{Introduction}
\label{sec:intro}

Galaxy outskirts, with their long dynamical times, retain the vestiges of their tumultuous formation history --- a history that is central to the now-standard, hierarchical picture of galaxy formation at the heart of the $\Lambda$CDM model \citep{blumenthal,davis};
a history that results in testable predictions for the total stellar debris field \citep[e.g.,][]{font, amorisco,merritt2}, the phase-space signatures of accretion events \citep{bullock,johnston,cooper,martin}, and the surviving population of gravitationally bound satellite galaxies \citep[][and references therein]{mao}; and a history that is manifested most clearly in the deepest possible images of galaxy halos \citep[surface brightnesses limits $>$ 30 mag arcsec$^{-2}$;][]{johnston}.

The hierarchical structure formation scenario successfully reproduces a myriad of observations, and is arguably one of the triumphs of cosmology, but comparison at galaxy scales and below can be problematic \citep{weinberg}. Interpreting the apparent failings is complicated by the complex details of the baryonic physics and its effects and by the unknown nature of dark matter. Nevertheless, a full accounting of the satellite mass function is now understood to provide a key test of the models, a possible path to learning more about both the effects of baryonic physics at these scales and the nature of dark matter, and motivation for numerous independent studies \citep[e.g.,][]{geha,park17,xi,park19,muller,carlsten,davisa,mao,k21,santos,nashimoto,li,goto}. 

Satellite galaxies represent the survivors of the accretion process, while the dispersed stars within the halos testify of the less fortunate. This interpretation is confirmed for the halo of our own galaxy, where $\gtrsim$ 95\% of all halo stars outside of surviving satellite galaxies are traced back to specific progenitor satellite galaxies \citep{naidu}. Among nearby galaxies there appears to be a wide variety in the amount of diffuse stellar debris that comprises the stellar halo \citep{sola}, including some galaxies with almost no stellar halo \citep{vdk}. Despite theoretical expectations of significant scatter in stellar halo properties \citep{cooper,martin}, the large 
variation observed may be in tension with the standard model \citep{merritt,merritt2}.

Our Large Binocular Telescope (LBT) Imaging of Galactic Halos and Tidal Structures (LIGHTS) survey is motivated by a desire to establish the range of stellar halo properties and test the basics of the galaxy formation scenario.
We aim to provide the best possible measurements of the diffuse halo stellar light for as large a sample of nearby galaxies as possible. Our focus is therefore on reaching as faint a surface brightness limit as possible, while retaining the highest possible angular resolution with which to classify and mask contaminating sources. 
We introduced the survey in \cite{LIGHTSs} by presenting and discussing the data for our first galaxy (NGC 1042). We demonstrated there how LIGHTS images could impact our understanding of stellar halos. 

Here we present our current sample of 25 galaxies and provide an overview of the sample selection and data processing, while also presenting results regarding the relation between the low surface brightness  satellite populations and their hosts. In short order, we will also be presenting a determination of the point spread function in the images to extremely large angular radii (Sedighi et al. in prep.) which is essential in assessing the scattered light across the image, and the generation of object catalogs (Hosseini-Shahisavandi et al. in prep.). Both of these are critical components in making the survey as scientifically useful as possible. As stressed in our introductory paper, LIGHTS is a well-matched precursor to the Legacy Survey of Space and Time (LSST) to be completed by the Vera C. Rubin observatory in somewhat over a decade, and so also serves as a guide on what to expect from that tremendous resource.

As just mentioned, our focus here is on low surface brightness (LSB) satellite candidates (central surface brightness in the $g$-band, $\mu_{0,g}$ $>$ 24 mag arcsec$^{-2}$). We refer to our cataloged satellite galaxies as candidates because they are not yet confirmed to be at the same distance as the apparent host. The surface brightness boundary is artificial, but does roughly correspond to the types of systems that are now being found in large numbers but were previously seen as rare \citep[e.g.,][]{vdk15a,koda}. We adapt the methodology used by \cite{smudges,smudges2,smudges3, smudges5}, which was specifically designed to identify LSB galaxies in Legacy Survey images \citep{dey} beyond the local flow field. 

The resulting catalog from LIGHTS images provides information on the nature of the satellite galaxy population at the limits of surface brightness and total luminosity that can be probed currently from the ground in galaxies well outside the Local Group and it complements Local Group or nearby galaxy studies by providing measurements for a larger set of parent galaxies. These results will eventually be matched to those on the diffuse halo light for these same galaxies to provide a complete accounting of the stellar halos of galaxies. 

In complementing our presentation here of the LIGHTS survey sample with some example results, we focus on two extremes of surface brightness properties. First, we discuss ultra-diffuse galaxies (UDGs), which are those LSB satellites that have effective radii larger than 1.5 kpc ($r_e > 1.5$ kpc). These galaxies are of interest because they include some of the most massive of the LSB galaxy population \citep{2017vanDokkum,zaritsky17}. 
Our satellite candidate catalog enables tests of previous results  that use different data, methodology, and parent galaxy type \citep[e.g.,][]{park17,carlsten,poulain,goto}.
For example, we will use our catalog to test the extrapolation of the relationship between the number of UDGs and parent host halo mass \citep{vdb,rt,mp,karunakaran2,goto}. Second, we discuss the incidence of high surface brightness central clusters (nuclear star clusters; NSCs) found in some LSB galaxies. 
The origin of NSCs remains in question \citep{neumayer} and their presence in LSB galaxies poses additional challenges to any formation scenario \citep{lambert}. Here we  explore
if there is evidence for a dependence on the incidence of NSCs in satellite galaxies with parent galaxy morphological type. 

This paper is structured as follows: in \S2, we present and describe the LIGHTS sample and the LIGHTS data reduction procedure through to the generation of final image mosaics. In \S\ref{subsec:image processing} we describe the modifications to the \cite{smudges5} pipeline, our recovery of LSB satellite galaxy candidates, and our estimation of the sample completeness. In \S\ref{sec:results} we present the basic properties of the candidate satellite sample, such as the number of satellites per parent and the number of UDGs. In \S\ref{sec:comparison} we compare the results to those of other studies. In \S\ref{sec:discussion} we discuss our results for the number of UDGs per host halo mass relative to the trend observed for more massive host halos and the subsample of nucleated LSB satellite candidates. Magnitudes are provided in the AB system \citep{oke1,oke2}.

\section{Sample and Data Reduction}
\label{sec:method}
\subsection{The Target Sample}

The LIGHTS project aims primarily to study the stellar halo structures of nearby disk galaxies. We set our target selection criteria for reasons based on the scientific aims of the survey, such as the ability to differentiate between extended stellar disk and halo components, and on pragmatic ones, such as the camera's field of view. We adopt the following selection criteria (with specific exceptions noted below): 1) declination $>-10^\circ$ to allow for relatively low airmass observations (airmass $<$ 1.4) from the LBT (at latitude $\sim$32$^\circ$); 2) an angular diameter that lies between 4 and 10 arcmin as measured by D$_{26}$, the diameter corresponding to the $r$-band 26 mag arcsec$^{-2}$ isophote, to reasonably match the field-of-view of the Large Binocular Cameras (LBC) instrument (see \S \ref{dataprocessing}); 3) an axis ratio, $b/a$ $ < 0.8$ to provide geometric contrast between a stellar halo and an extended disk component; 4) a Galactic reddening,  E$(B-V)$, $< 0.04$ as measured by \cite{SFD} to avoid regions of high Galactic extinction; 5) a redshift, $z$, less than $0.006$ (closer than $\sim26$ Mpc) to limit the study to nearby galaxies and provide us with high physical resolution (corresponding to $\sim$ 30 pc pixel$^{-1}$ at the native image resolution and 100 pc pixel$^{-1}$ in the binned images we will use for detecting low surface brightness satellite candidates; \S\ref{subsec:image processing}); and 6) an absence of nearby projected bright stars (first determined using an automated pass through the Yale Bright Star Catalog \citep{BSC} to search for stars within 1$^\circ$ and then a visual examination based on experience as the program progressed). The data we use to make this selection (Table \ref{tab:targets}) are drawn from the Sienna Galaxy Catalog 2020 \citep{moustakas},
which provides photometric measurements of galaxies of large angular extent from the Legacy Survey data \citep{dey}, and ancillary data drawn from the Hyperleda Extragalctic Database\footnote{\url{http://leda.univ-lyon1.fr/}} \citep{makarov}, except for the distances (see below),
and the number of identified LSB candidate satellites, N$_{LSB}$, and ultra-diffuse galaxies, N$_{UDG}$, which
we measure here. The number of galaxies projected within 100 kpc and having a measured recessional velocity within 500 km s$^{-1}$ of the target galaxy in the Hyperleda database are tabulated as N$_{LEDA}$. The radial cut matches the projected radii probed by the LIGHTS images and the velocity cut is set to include both bound satellites and potential group members.

We adopt published distances from  
studies utilizing standard candles (SNe, TRGB, and Cepheids), scaling relations (Tully-Fisher), and detailed cosmological flow modeling. We chose which distance measurements to adopt based on our own evaluation of the relative merits and the selected references are noted in  Table \ref{tab:targets} as D$_{ref}$.
Distance estimates, particularly at distances where the peculiar velocity is not $\ll$ than the Hubble flow velocity, can be highly uncertain. In extreme cases, literature studies can disagree on the distance by as much as a factor of two. The notable example within our sample is NGC 1042, for which our adopted distance of 13.5$\pm$2.6 Mpc is markedly different than that presented in a separate study \citep[20.0$\pm$1.6 Mpc;][]{danieli}. Although this difference has significant repercussions on certain questions of great interest \cite[e.g.,][]{trujillo19,danieli} and the choice of distance impacts the physical parameters we present for our satellite candidates, it does not significantly affect the results presented here from our overall sample. Nevertheless, uncertainties in the derived physical parameters of our candidates, even for galaxies other than NGC 1042, are dominated by distance uncertainties.

We violate our selection criteria on occasion, as is evident in Table \ref{tab:targets}. Specifically, we exceed the D$_{26}$ criterion by including NGC 2903, NGC 5033, and NGC 5907, which all have D$_{26}>10$ arcmin (12.82, 11.53, and 21.16 arcmin, respectively). The angular sizes of these make the images more difficult to reduce, even when our dithering steps are larger. Two of these are included because
they are part of an earlier leading study of stellar halos \cite[NGC 2903 and NGC 5907;][]{merritt}, and so are valuable for comparison.
We also violate the $b/a < 0.8$ criterion by including NGC 1042, NGC 3351, NGC 3486, and NGC 3596 (0.87, 0.89, 0.87, and 0.91, respectively). Two of these (NGC 1042 and NGC 3351) are included because they are also in the \cite{merritt} study. Regarding the remaining galaxies that violate the criteria and are not in an earlier study, we  included them due to a shortage of more suitable targets. 

\begin{longrotatetable}
\begin{deluxetable}{lrrrrrrrrrrrrrrr}
\tablecaption{LIGHTS Target Galaxies$^a$}
\label{tab:targets} 
\tablehead{ 
\colhead{Name}&
\colhead{RA}&
\colhead{Dec}&
\colhead{$m_{g}$}&
\colhead{$m_{r}$}&
\colhead{D}&
\colhead{$\sigma_{\rm D}$}&
\colhead{D26}&
\colhead{E(B-V)}&
\colhead{b/a}&
\colhead{z}&
\colhead{Morph}&
\colhead{N$_{LSB}$}&
\colhead{N$_{UDG}$}&
\colhead{N$_{LEDA}$}&
\colhead{D$_{ref}$}\\ 
&
&
&
\colhead{[mag]}&
\colhead{[mag]}&
\colhead{[Mpc]}&
\colhead{[Mpc]}&
\colhead{[$^\prime$]}}
\startdata
           NGC1042 &    40.09986 &    $-$8.43354 & 11.64 & 11.07 &  13.5 &   2.6 &   5.47 & 0.029 &  0.87 & 0.0046 &       SABc &    7 &    0 &    6 &     1\\
           NGC2712 &   134.87698 &    44.91390 & 12.33 & 11.62 &  30.2 &   2.0 &   4.49 & 0.020 &  0.54 & 0.0061 &        SBb &    2 &    0 &    0 &     2\\
           NGC2903 &   143.04212 &    21.50083 &  9.52 &  8.81 &  10.0 &   2.5 &  12.82 & 0.031 &  0.52 & 0.0019 &        Sbc &    0 &    0 &    1 & 2+3+4\\
           NGC3026 &   147.73071 &    28.55111 & 13.32 & 12.79 &  15.8 &   2.0 &   4.64 & 0.021 &  0.26 & 0.0049 &        SBm &    0 &    0 &    0 &     5\\
           NGC3049 &   148.70652 &     9.27109 & 12.95 & 12.31 &  19.3 &   0.2 &   4.34 & 0.038 &  0.50 & 0.0050 &        SBb &    0 &    0 &    1 &     3\\
           NGC3198 &   154.97897 &    45.54962 & 10.95 & 10.36 &  12.9 &   0.1 &   9.57 & 0.120 &  0.36 & 0.0022 &         Sc &    0 &    0 &    1 &     3\\
           NGC3351 &   160.99042 &    11.70381 & 10.29 &  9.53 &  10.0 &   0.3 &   7.58 & 0.028 &  0.89 & 0.0026 &         Sb &    0 &    0 &    1 &     6\\
           NGC3368 &   161.69058 &    11.81994 &  9.79 &  9.02 &  11.2 &   0.5 &   9.54 & 0.025 &  0.75 & 0.0030 &        Sab &    4 &    0 &    0 &     6\\
           NGC3486 &   165.09945 &    28.97514 & 11.00 & 10.42 &  13.6 &   0.1 &   6.30 & 0.022 &  0.87 & 0.0023 &         Sc &    1 &    0 &    0 &     4\\
           NGC3596 &   168.77586 &    14.78702 & 11.85 & 11.28 &  11.3 &   1.1 &   3.99 & 0.024 &  0.91 & 0.0040 &       SABc &    1 &    0 &    2 &     6\\
           NGC3675 &   171.53575 &    43.58592 & 10.29 &  9.42 &  16.8 &   2.0 &   7.38 & 0.020 &  0.47 & 0.0026 &         Sb &    2 &    0 &    0 &     2\\
           NGC3726 &   173.33802 &    47.02920 & 10.74 & 10.11 &  13.6 &   2.0 &   8.39 & 0.017 &  0.60 & 0.0029 &         Sc &    1 &    0 &    0 &     2\\
           NGC3941 &   178.23066 &    36.98633 & 10.88 & 10.19 &  13.8 &   0.1 &   4.90 & 0.021 &  0.68 & 0.0030 &         S0 &    1 &    0 &    0 &     3\\
           NGC3953 &   178.45382 &    52.32677 & 10.65 &  9.82 &  18.8 &   1.0 &   8.32 & 0.030 &  0.50 & 0.0035 &        Sbc &    0 &    0 &    2 &     7\\
           NGC3972 &   178.93787 &    55.32074 & 12.27 & 11.60 &  20.8 &   0.2 &   8.27 & 0.014 &  0.27 & 0.0028 &       SABb &    3 &    0 &    9 &     8\\
           NGC4010 &   179.65787 &    47.26150 & 12.81 & 12.09 &  17.9 &   2.0 &   7.80 & 0.024 &  0.22 & 0.0030 &       SBcd &    1 &    0 &    0 &     2\\
           NGC4220 &   184.04880 &    47.88326 & 11.78 & 10.94 &  20.3 &   2.0 &   5.98 & 0.018 &  0.33 & 0.0031 &       S0-a &    3 &    1 &    2 &     9\\
           NGC4307 &   185.52368 &     9.04363 & 12.05 & 11.23 &  20.0 &   2.0 &   7.77 & 0.023 &  0.22 & 0.0035 &        SBb &    5 &    2 &    5 &     2\\
           NGC4321 &   185.72846 &    15.82182 & 10.21 &  9.53 &  15.2 &   0.5 &   8.63 & 0.026 &  0.84 & 0.0053 &       SABb &    4 &    0 &    4 &    10\\
           NGC4689 &   191.93985 &    13.76281 & 11.38 & 10.73 &  15.0 &   2.2 &   6.13 & 0.023 &  0.77 & 0.0054 &         Sc &    2 &    0 &    0 & 2+3+4\\
           NGC5033 &   198.36444 &    36.59394 & 10.68 &  9.93 &  19.1 &   2.0 &  11.35 & 0.012 &  0.44 & 0.0029 &         Sc &    2 &    0 &    0 &     9\\
           NGC5248 &   204.38343 &     8.88518 & 10.89 & 10.16 &  14.9 &   1.3 &   7.54 & 0.024 &  0.67 & 0.0038 &       SABb &    2 &    0 &    0 &    11\\
           NGC5866 &   226.62291 &    55.76321 & ... & ... &  14.1 &   0.5 &   7.89 & 0.013 &  0.40 & 0.0022 &       S0-a &    6 &    1 &    2 &    12\\
           NGC5907 &   228.97404 &    56.32877 & 10.50 &  9.66 &  16.5 &   0.1 &  21.16 & 0.011 &  0.11 & 0.0033 &       SABc &    6 &    0 &    2 &     3\\
           NGC6015 &   237.85512 &    62.31003 & 11.16 & 10.54 &  19.0 &   0.2 &   9.10 & 0.013 &  0.48 & 0.0028 &         Sc &    1 &    0 &    0 &     3\\
\enddata
\tablenotetext{}{\scriptsize Reference codes (D$_{ref}$) correspond to: (1) \cite{monelli};  (2) \cite{tully2016}; (3) \cite{shaya}; (4) \cite{kourkchi};  (5) \cite{nasonova}; (6) \cite{jacobs};  (7) \cite{bose}; (8) \cite{riess}; (9) \cite{tully2013}; (10) \cite{freedman}; (11) \cite{kourkchi2017};  (12) \cite{cantiello}}
\end{deluxetable}
\end{longrotatetable}

\begin{figure}[ht]
    \centering
    \includegraphics[scale=0.45]{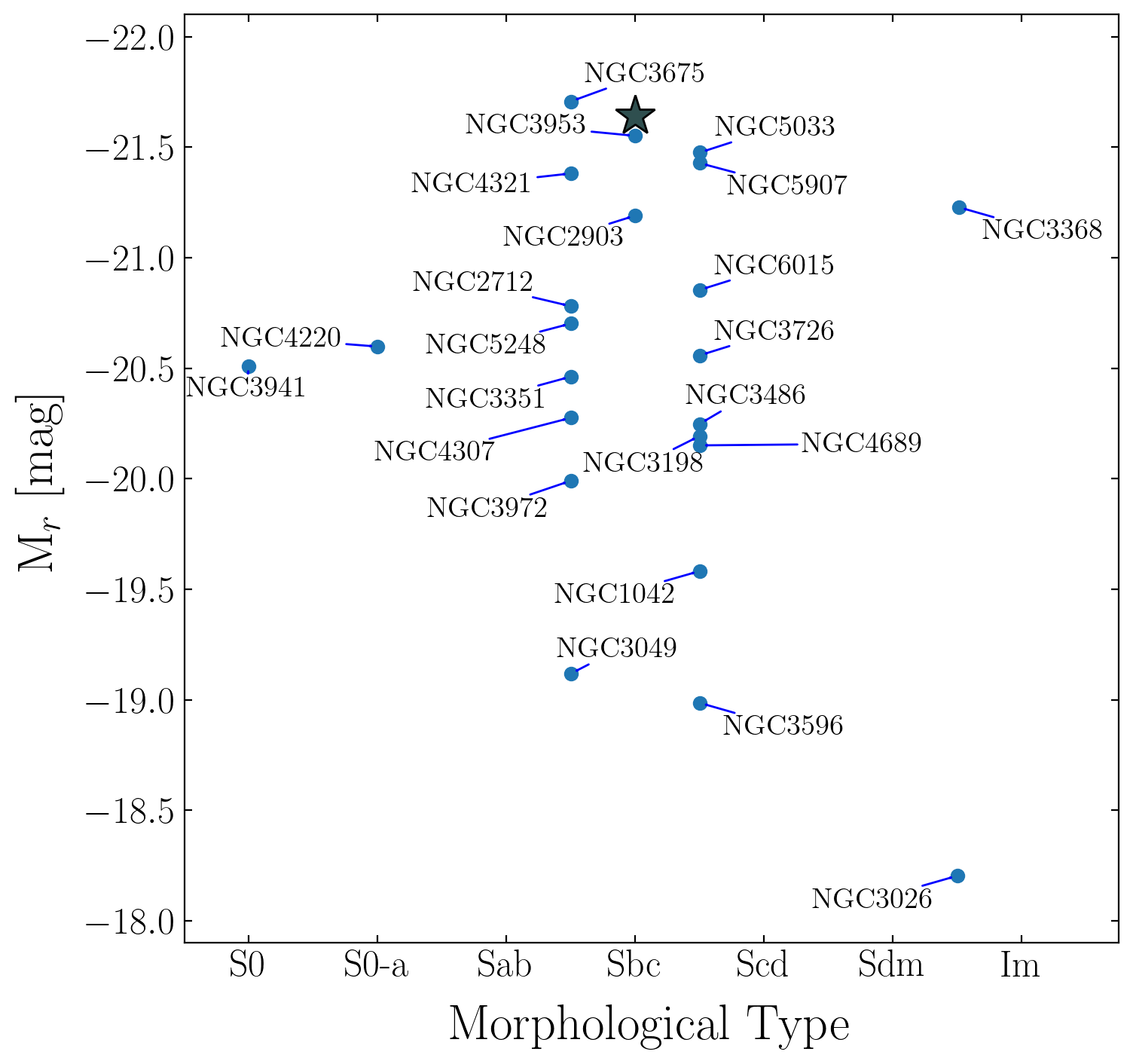}
    \caption{The LIGHTS sample to date in the morphology-absolute magnitude space. The star represents the Milky Way for comparison, with an adopted absolute magnitude of M$_r=-21.64$ \citep{bland-hawthorn} and a morphological classification of Sbc \citep{hodge,kennicutt01}. 
    }
    \label{fig:sample}
\end{figure}

In Figure \ref{fig:sample} we show that the sample consists of galaxies that are morphologically similar to the Milky Way, but have a median magnitude that is $\sim$ 1 mag fainter ($-20.6$). The sample, as a whole, cannot be characterized as Milky Way analogs, although a subsample of $\sim$ 5 galaxies (NGC 3675, NGC 3953, NGC 4321, NGC 5907, and NGC 5033) can be defined as such if one desires a closer comparison to the MW satellite system.

\begin{rotate}
\begin{deluxetable*}{cccccccccccccc}
\tablecaption{LIGHTS Target Galaxies: Observational Details}
\label{tab:giulia}
\tablehead{
\colhead{Name}&
\colhead{FWHM$_g$\tablenotemark{a}}&
\colhead{FWHM$_r$\tablenotemark{b}}&
\colhead{$<$airmass$>$}&
\colhead{FOV\tablenotemark{c}}&
\colhead{FOV(70$\%$)\tablenotemark{d}}&
\colhead{T$_{exp}$}&
\colhead{$<$T$_{exp,70}$$>$}&
\colhead{$\mu_{lim,g}$\tablenotemark{e}}&
\colhead{$m_{lim,g}$\tablenotemark{f}}&
\colhead{$\mu_{lim_r}$}&
\colhead{$m_{lim,r}$}
\\
&
\colhead{[arcsec]}&
\colhead{[arcsec]}&
\colhead{}&
\colhead{[arcmin]}&
\colhead{[arcmin]}&
\colhead{[hr]}&
\colhead{[hr]}&
\colhead{[mag arcsec$^{-2}$]}&
\colhead{[mag]}&
\colhead{[mag arcsec$^{-2}$]} &
\colhead{[mag]}
}
\startdata
NGC 1042 & 0.99 & 1.00 & 1.38 & 37.0 & 20.0 & 1.5 & 1.27 & 31.53& 27.77 & 30.69 & 26.83 \\
NGC 2712 & 1.03 & 1.00 & 1.24 & 37.7 & 18.4 & 1.5 & 1.29 & 31.83 & 28.22 & 30.85 & 27.06\\
NGC 2903 & 1.19 & 1.21 & 1.18 & 49.1 & 19.0 & 1.5 & 1.17 & 30.80 & 26.89 & 30.26 & 26.14\\
NGC 3026 & 1.30 & 1.55 & 1.17 & 39.2 & 22.3 & 1.5 & 1.34 & 30.57 & 26.61 & 30.00 & 25.83\\
NGC 3049 & 1.09 & 1.10 & 1.10 & 38.4 & 21.3 & 1.0 & 0.85 & 31.14 & 27.43 & 30.45 & 26.71\\
NGC 3198 & 1.37 & 1.44 & 1.16 & 42.4 & 22.1 & 1.5 & 1.33 & 31.42 & 27.38 & 30.65 & 26.60\\
NGC 3351 & 2.05 & 2.02 & 1.29 & 42.3 & 16.6 & 1.0 & 0.84 & 30.87 & 26.43 & 30.18 & 25.72\\
NGC 3368 & 1.12 & 1.16 & 1.11 & 39.4 & 20.7  & 1.5 & 1.34 & 31.27 & 27.42 & 30.40 & 26.58\\
NGC 3486 & 1.60 & 1.59 & 1.07 & 40.2 & 18.1 & 1.5 & 1.21 & 31.36 & 27.10 & 30.54 & 26.38\\
NGC 3596 & 1.30 & 1.20 & 1.08 & 37.8 & 22.1 & 1.6 & 1.43 & 31.50 & 26.74 & 30.84 & 26.12\\
NGC 3675 & 0.91 & 1.17 & 1.03 & 41.9 & 21.0 & 1.5 & 1.33 & 31.34 & 27.71 & 30.64 & 26.58\\
NGC 3726 & 1.10 & 1.25 & 1.34 & 42.0 & 21.1 & 1.5 & 1.32 & 31.10 & 27.28 & 30.69 & 26.83\\
NGC 3941 & 1.00 & 1.22 & 1.32 & 40.7 & 22.2 & 1.45 & 1.30 & 31.43 & 27.79 & 30.75 & 26.81\\
NGC 3953 & 1.23 & 1.40 & 1.14 & 42.2 & 21.0 & 1.5 & 1.32 & 31.22 & 27.44 & 30.69 & 26.56\\
NGC 3972 & 1.05 & 1.22 & 1.14 & 43.9 & 20.6 & 1.5 & 1.21 & 31.23 & 27.43 & 30.60 & 26.75\\
NGC 4010 & 1.99 & 1.52 & 1.05 & 37.5 & 22.8 & 0.5 & 0.42 & 30.38 & 25.95 & 29.88 & 25.72\\
NGC 4220 & 1.85 & 1.69 & 1.27 & 43.9 & 22.2 & 1.75 & 1.58 & 31.52 & 27.10 & 30.96 & 26.68\\
NGC 4307 & 1.17 & 1.69 & 1.15 & 37.5 & 20.8 & 2.5 & 2.23 & 31.56 & 27.70  & 30.92 & 26.84\\
NGC 4321 & 1.14 & 1.25 & 1.23 & 41.5 & 19.3 & 2.0 & 1.74 & 30.70 & 26.91 & 29.93 & 26.13\\
NGC 4689 & 1.61 & 1.64 & 1.29 & 37.7 & 21.0 & 1.0 & 0.85 & 31.01 & 26.81 & 30.39 & 26.13\\
NGC 5033 & 1.15 & 1.33 & 1.20 & 44.8 & 18.8 & 1.7 & 1.48 & 31.35 & 27.41& 30.56 & 26.55\\
NGC 5248 & 1.05 & 1.03 & 1.13 & 39.1 & 20.6 & 1.45 & 1.29 & 31.23 & 27.61 & 30.60 & 26.89\\
NGC 5866 & 1.14 & 1.24 & 1.09 & 41.7 & 20.4 & 3.1 & 2.77 & 32.01 & 28.24 & 31.20 & 27.13\\
NGC 5907 & 1.19 & 1.35 & 1.10 & 50.2 & 20.4 & 1.5 & 1.28 & 31.44 & 27.45 & 30.80 & 26.68\\
NGC 6015 & 1.24 & 1.34 & 1.22 & 44.7 & 22.3 & 1.5 & 1.34 & 31.29 & 27.30 & 30.70 & 26.62\\                                                                                                                                                                                                                                                                                                                    
\enddata
\tablenotetext{a}{Measured in $g$-band using point-like sources in the final image.}
\tablenotetext{b}{Measured in $r$-band using point-like sources in the final image.}
\tablenotetext{c}{The final field-of-view (FOV) is a square with a length of the given value.}
\tablenotetext{d}{The FOV(70$\%$) is the field observed for T$_{exp}$ $>$ 70$\%$.
The FOV(70$\%$) is a circle with a diameter of the given value.}
\tablenotetext{e}{Surface brightness limit is computed using the extended metric of 3$\sigma$ detection in areas equivalent to 10 arcsec by 10 arcsec boxes.
}
\tablenotetext{f}{Limiting magnitude for point-like sources computed within a FWHM radius circular aperture (5$\sigma$ detections).}
\end{deluxetable*}
\end{rotate}

\subsection{LIGHTS Processing Overview}
\label{dataprocessing}

We obtained ultra-deep observations of LIGHTS targets using the LBT and both channels (blue and red) of the LBC simultaneously \citep{lbc}. The observations presented here were carried out during Director Discretionary Time (DDT, P.I. D'Onofrio) and standard allocated time (P.I. Zaritsky) beginning in Oct 2020 through the first half of 2023.
Each of the two LBC cameras consist of 4 CCDs,  with a pixel scale of 0.224 $\arcsec$ pixel$^{-1}$, covering an approximate area of 7.8$\arcmin$ $\times$ 17.6$\arcmin$ per CCD, with $\sim$ 18 $\arcsec$ wide gaps between the chips. The combined field of view for each LBC camera is roughly 23$\arcmin$ $\times$ 25$\arcmin$. We used the g-SLOAN filter on the LBC Blue camera and the r-SLOAN filter on the LBC Red camera optimized for the wavelength ranges 3500 \AA\ to 6500 \AA\ and 5500 \AA\ to 1 $\mu$m, respectively.

Attentive observing and data reduction procedures are required to address different and
complex observational challenges such as scattered light, strongly saturated stars,
and ghosts. Furthermore, to obtain images in which low surface brightness features are reliable (i.e. limiting surface brightness of  $\mu_g$ $\sim$ 31 mag/arcsec$^2$, 3$\sigma$ in a $10\arcsec \times 10\arcsec$ box), 
accurate flat field estimation and careful treatment of the sky
background are integral. The latter task in particular is challenging because observational conditions, like the air mass and light cloud coverage, change throughout the night. To mitigate these issues, we apply a specific observational strategy involving a dithering pattern using ten 180 sec exposures per sequence \citep{LIGHTSs}, with which the total exposure time is built up, and an optimized data reduction pipeline based on extensive prior experience \citep[e.g.,][]{tf}. 
The data reduction pipeline relies on GNU Astronomy Utilities \citep[Gnuastro;][]{gnuastro,gnuastro2}.
Below, we provide an overview of the key stages in this process. Further details will be provided by Golini et al. (in prep.).

\subsubsection{Data Preprocessing and Bias Subtraction}

At this initial stage, we work on each CCD of the camera individually.
To prepare the data for processing, we create a bad pixel mask that covers the pixels that do not contribute to the signal due to detector readout failures and where saturation is leading to a non-linear detector response. We subtract the overscan, crop
the overscan region, construct a master bias image for each CCD, and subtract the corresponding master bias from each image.
The master bias is constructed by combining 30 individual bias frames using a sigma clip mean (3$\sigma$) algorithm provided in  Gnuastro's Arithmetic program (\texttt{astarithmetic}).

\subsubsection{Flat field Correction}

Precise flat field correction is crucial for our research.
Dome flats are unsuitable for deep imaging due to irregularities in the LBT dome illumination. 
Moreover, twilight flats do not fully meet our needs due to variations between night-sky illumination and the twilight light gradient from the horizon where the Sun has set.
To sidestep these issues, we create a master flat using our own set of science images. Although one might suspect that using the complete set of science images would produce a higher-quality master flat, small variations in focus, vignetting, airmass, and moon illumination from night to night, along with light pollution from nearby populations and the scattered light from bright stars,  result in less than optimal flat field frames.
To address this problem, we use only data obtained on each night to create the flat field image for that night's data. 
We require at least 15 science images each night with which to create the master flat for that specific night. This minimum number is essential for statistical robustness.

For every night of observation,  each filter, and each CCD, we generate a distinct master flat using bias-corrected science images. 
Our process involves two steps. 
First, we identify 
those CCD images that we need to reject because the large-scale illumination is grossly non-uniform. Such images deviate from the typical background illumination, potentially introducing erroneous gradients that become noticeable in the final stack. We have a variety of criteria that we impose. We reject
CCD images that contain a bright star (M$_{{\rm V}}<9$ mag) or where the target galaxy covers 60$\%$ or more of the CCD. 
We also reject from the stack any CCD images where the distance between the central pixel and a bright star is small (a sliding scale based on the brightness of the star being considered, ranging from 7 arcmin for a 5.5 mag star to 5 arcmin for an 8.5 mag star) or where 
the distance between the central pixel and the target galaxy is equal to or less than half the semi-major axis of the galaxy.
The percentage of images excluded per CCD varies depending on the specific field, ranging from 5$\%$ to 30$\%$ in both $g$ and $r$ filters.
By employing dithering with large displacements, we ensure that there are enough frames that are clear of the central galaxy for every CCD. 
Then, we create the master flat using the procedure described by \cite{LIGHTSs}.
Bias-corrected science images in each filter are divided by their corresponding CCD final flat field image.
To avoid vignetted regions at the corner of the detectors, we remove from the images all pixels where the final flat field image has an illumination fraction that is $<$ 0.9.

\subsubsection{Astrometry, Sky Subtraction, and Photometric Calibration}
\label{subsub:skysub}
We determine the astrometry of our individual science images as follows. 
We initially compute the astrometric solution using Astrometry.net (v0.85; \citealt{astrometry.net}) and use Gaia eDR3 \citep{gaia} as our astrometric reference catalog.
Because each CCD of the LBC exhibits its own distortions, we employ SExtractor \citep[v.2.25.2;][]{sextractor} to create object catalogs for each CCD and SCAMP \citep[v.2.10.0;][]{scamp2006}, again using Gaia eDR3 as the reference catalog, to obtain higher fidelity astrometric solutions.
Using the astrometric solutions obtained using SCAMP, we next apply SWarp \citep{Swarp} to each individual image to map it to a common coordinate grid.

Proper sky subtraction is key when dealing with low surface brightness data to avoid introducing artificial gradients and impacting the subtle structures that might be there.
We assume that each individual CCD image's sky level can be approximated by a single constant value. We do this to avoid inadvertently modeling out physical low surface brightness features and expect our extensive dithering to mitigate spatial variations in the sky. Gradients due to scattered light from bright stars will be addressed separately using a well-determined point spread function. To compute the sky, we employ the $\texttt{-{}-checksky}$ option in NoiseChisel \citep{gnuastro2}. 
Initially, we mask bright sources and diffuse light because including them will result in a background value that is biased high and result in an over-subtraction of the background. 
Subsequently, we determine the sky value for each image by calculating the 3$\sigma$ clipped median of the non-masked pixels.
By working with CCD images covering an area of approximately 14$\arcmin$ $\times$ 8$\arcmin$, we ensure that we obtain a locally representative sky value for that specific region. Subsequently, for each CCD image taken throughout the night of observation, we subtract the corresponding sky value.

After sky subtraction, we convert the LBT counts (ADUs) into nanomaggies\footnote{see \url{https://www.sdss3.org/dr8/algorithms/magnitudes.php}}.
To do this, for each set of images, we construct a catalog of sources in common between the SDSS and LBT images, cross-referencing with Gaia eDR3 to confirm we have point-like sources, and measure fluxes within circular apertures with a diameter of 2$^{\prime\prime}$ \citep[for Gnuastro routines to do this see][]{Eskandarlou_2023}. 
We compute the flux ratios between the SDSS and LBT sources for each CCD separately.
We calculate the resistant (3$\sigma$ rejection) median of these ratios, and multiply the LIGHTS image by this median to obtain the pixel values expressed in units of nanomaggies.
When working with images that may exhibit variations in signal, noise, or instrumental effects, performing astrometric and photometric calibration before stacking allows for optimal control. Comparing across images with different image quality and using apertures of different sizes, we conclude that the zero point uncertainties are in the range of 0.01 to 0.02 mag.

\subsubsection{Image Stacking}

We combine individual exposures to create a final mosaic image using a weighted average
because observational conditions change throughout the night. Some of these variations are well understood and to some degree under our control, such as those that depend on the air mass and depend on the position of the target in the sky, and some are not well understood and beyond our control, such as meteorological events like passing clouds that affect the sky brightness. The latter will generate artificially high standard deviations in the sky pixel values, indicative of poorer image quality.
To mitigate these variations, images obtained in better conditions should carry more weight in the final mosaic. Therefore, we use a weighted average to combine the images, with a weight for the i$^{th}$ image equal to the ratio of the standard deviation of the sky in the best exposure to the standard deviation of the sky in the i$^{th}$ image. 

To compute the standard deviation of the sky in each image we begin by using the procedure described in \S. \ref{subsub:skysub} to create a mask for bright sources and diffuse light.
We then calculate the 3$\sigma$ clipped standard deviation of the non-masked pixels in each image.
A complication in calculating the weighted mean is that it is strongly influenced by outliers such as cosmic rays.
Cosmic rays typically manifest as isolated pixels with high values (exceeding 3$\sigma$ from the mean) and are correctly not recognized as sources by NoiseChisel.
We identify such pixels as those having a value that deviates from the sigma-clipped mean (3$\sigma$ rejection is sufficient for our purpose) of all pixels at the same sky location, in all images.
We mask these outliers and stack the data using the \texttt{"sigclip"}-weighted mean routine in Gnuastro's Arithmetic program. 
The co-added image is significantly deeper than any individual image, and therefore, low surface brightness features, previously invisible, emerge from the noise.
Therefore, we repeat the sky estimation (and subsequent reduction steps described previously) on the individual images using the improved masks generated by the first data co-add to produce our final data co-add.
We provide a detailed quantification of each step of the data reduction pipeline in
Golini et al. (in prep.) for the specific case of NGC 3486.

We tabulate the observational parameters, final image quality, and limiting magnitudes in Table \ref{tab:giulia}. 
The quoted limiting magnitude and surface brightness limits correspond to regions within each field that have been observed for more than 70$\%$ of the total exposure time. In calculating the surface brightness limits, we apply  the mask created from the deeper $g$ + $r$ data. We evaluate the variance in the unmasked pixel values for each image, adopt that as representative of the uncertainty in each pixel, and then calculate the 5$\sigma$ limit for a circular aperture with a radius equal to that of FWHM of each particular image to estimate the limiting magnitude and the 3$\sigma$ limit for an area equivalent to a 10 by 10 arcsec box to estimate the limiting surface brightness. As such, we assume that each pixel is independent.
We find a range of approximately one magnitude in our limits across fields that have the same exposure time. We have traced some of this variation to differences in seeing conditions, air mass, scattered light, light pollution, and proximity to the moon and its illumination.
The moon has the clearest effect in the magnitude limit, although a large fraction of the total variation in the limiting magnitude remains unaccounted for and presumably reflects variations in the atmospheric properties at the time of observation.

\section{Our LSB Satellite Search}
\label{subsec:image processing}

We base our pipeline for identifying LSB satellite galaxy candidates on that used for our processing of the Legacy Survey \citep{smudges5}, but major modifications are required for LIGHTS because of its much greater depth, larger file sizes and the availability of only $g$ and $r$ bands \citep[the Legacy Survey also has $z$ band observations;][]{dey}.  The major steps for identifying potential LSB satellites in LIGHTS are: 1) image processing that produces a preliminary list of candidates;  2) visual confirmation of remaining candidates; 3) estimation of completeness using simulated sources; and 4) creation of the catalog. 

This procedure is a significant departure from \cite{smudges5} where we also rejected candidates based on thresholds for cirrus contamination and screened survivors with a machine-learning classifier before visual confirmation. These steps are unnecessary in LIGHTS because target galaxies were already selected in regions of low extinction and its relatively small total footprint, as opposed to the $\sim$20,000 deg$^2$ of the Legacy Survey, allows for visual confirmation of all candidates.  

Details of the Legacy Survey pipeline have been previously described and, other than brief summaries, we only address modifications here.  Because of the reasons noted in \S \ref{subsec:image processing}, Step 3, we make two passes through the pipeline and the number of potential candidates surviving each step discussed below represents the final total after the two runs.  Processing and analysis of calibrated LIGHTS images is performed using the facilities of the University of Arizona High Performance Computing center\footnote{\href{https://public.confluence.arizona.edu/display/UAHPC/Resources}{public.confluence.arizona.edu/display/UAHPC/Resources}}. 

We identify potential LSB satellite galaxy candidates using the following steps and summarize these in Table \ref{tab:screening}: 
\begin{enumerate}

\item Because of the large size of the LIGHTS mosaics, we start by binning them by a factor of four in each direction, which also increases the signal-to-noise ratio.  While this increases the pixel scale from 0.224 arcsec pixel$^{-1}$ 
to 0.896 arcsec pixel$^{-1}$, the effective pixel size is still significantly smaller than our minimum allowable candidate satellite size (3.5 rebinned pixels or $r_e = 3.2 \arcsec$, which we find to be a coarse lower size limit on the resolution needed for reliable model fitting and classification). The LIGHTS pipeline flags saturated regions, bad pixels, and image regions outside of the observation area with NaN values.  We replace these NaN values using the Python ASTROPY interpolate\_replace\_nans function.

\item To limit the number of candidates requiring further processing in the Legacy Survey, \cite{smudges5} subtracted sources that were clearly too bright to be their intended targets. This step required masking out to the periphery of these bright objects.  However, the significantly greater depth of the LIGHTS images results in a much higher density of such objects and we skip this step to avoid excessive masking.

\item As described in detail by \cite{smudges}, the detection pipeline uses wavelet transforms with tailored filters to isolate candidates of different angular scales.  Higher wavelet levels will preferentially accentuate objects of larger sizes. However, clusters of small objects can be detected across multiple wavelet levels.  To limit the number of false detections when processing Legacy Survey images, \cite{smudges5} required that higher order wavelets have a peak value of at least 25\% of lower order wavelets within its footprint.  Because of the greater depth in LIGHTS, and the resulting larger number of overlying contaminants, we found that we initially missed some obvious LSB candidates when using this criterion and, therefore, chose to repeat our processing without this requirement. 
With this change, we find a total of 717,124 wavelet detections.

\item  We limit spurious detections by requiring that a potential candidate have coincident detections in both filter bands (defined as center-to-center separations $<$ 4 arcsec) with its resulting location defined as the midpoint of the two centers. This criterion results in the acceptance of 541,086 wavelet detections which are, after eliminating nearby duplicates (defined to be other detections within 4 arcsec),  associated with 261,046 groups of detections needing further processing.

\item
The vast majority of detections surviving the previous steps are still not viable LSB satellite galaxy candidates and require further filtering. Rather than using time-consuming GALFIT modeling \citep{peng} at this point, we obtain rough parameter estimates by fitting an exponential S\'ersic model ($n=1$) with the much faster LEASTSQ function from the Python SciPy library \citep{jones}. Because this modeling is only used as a coarse selection screen, we require that results meet conservative parameter thresholds of
$r_e \ge 2.7^{\prime\prime}$ and $\mu_0 \ge$  23.0 and 22.0 mag arcsec$^{-2}$ for $g$ and $r$ bands, respectively, to avoid prematurely rejecting good candidates. An important clarification, particularly with respect to our discussion of nucleated sources in \S\ref{sec:nucleated}, is that we mask bright regions in the center of each candidate when fitting models \citep[see][for details]{smudges5}. Despite this initial use of  a fixed $n$, final samples from this procedure show no bias against higher $n$ objects at least up to $n = 2$ \citep{smudges2}. A total of 211,104 detections comprising 153,552 distinct candidates survive this step. 

\item We now revisit the modeling using the combination of the $g$ and $r$ band images to reach greater depth, a slightly refined model, and additional selection criteria. We now do use GALFIT to model each candidate using a fixed S\'ersic index, $n=1$, and again use generous acceptance thresholds of $r_e \ge 2.7\arcsec$, $b/a \ge 0.34$,  and $\mu_{0,g} \ge$ 23 mag arcsec$^{-2}$ or $\mu_{0,r} \ge$ 22.2 mag arcsec$^{-2}$ if there is no available measurement of $\mu_{0,g}$. Once again the values are set to avoid prematurely rejecting acceptable candidates. A total of 27,209 candidates meet these criteria. 

\item  In our final image processing step, we model the remaining candidates using GALFIT with a variable S\'ersic index and an estimate of the PSF derived from the FWHM parameter given in Table \ref{tab:giulia} for both $g$ and $r$-band images. To avoid having a nuclear star cluster distort the fit, we mask the center if those pixels are a factor of two brighter than our central surface brightness limits (24 and 23.6 mag arcsec$^{-2}$ for $g$ and $r$, respectively). After applying our final, tighter LSB candidate criteria of $r_e \ge 3.2\arcsec$, $\mu_{0,g}\ge$ 24 mag arcsec$^{-2}$ (or $\mu_{0,r}\ge$ 23 mag arcsec$^{-2}$ if GALFIT failed to model the $g$-band image), $b/a \ge $ 0.37, and $n < 2$ we are left with a total of 6044 candidates requiring further evaluation.
These parameter choices mirror those in \cite{smudges5} and are discussed in more detail there.

\item The remaining candidates are visually reviewed by DZ and RD with each candidate labeled as a potential LSB satellite or a false positive. Those with disagreements are classified again by both reviewers.  This procedure results in both reviewers labeling 62 as LSB candidates with the remainder classified as a false positive by at least one of the reviewers. To minimize the number of false candidates in our final list, we consider any disagreements between the two reviewers to be false positive (5 objects are in this category). 
\end{enumerate}

Visual classification, in fact any type of morphological classification, has the potential to introduce unintended biases in the properties of the final sample. We attempt to reject only objects that are evidently not galaxies (tidal tails, merged groups of background galaxies, image artifacts) but biases could nevertheless creep in. In our comparison of the SMUDGes catalog with other catalogs \citep{smudges2,smudges5} and 
our comparison here to independent work (\S 5) we do not identify any significant classification disagreements among galaxies in common that satisfy the stated selection criteria of either survey, suggesting that we are not introducing any peculiar biases with our visual classification, although it is possible that all surveys share similar biases.

\begin{deluxetable*}{lcrr}
\tablecaption{Number of Detections and LSB Satellite Galaxy Candidates in LIGHTS vs. Processing Step \tablenotemark{a} }
\label{tab:screening}
\tablewidth{0pt}
\tablehead{
\colhead{Process}&Step in \S\ref{subsec:image processing} &
\colhead{Detections\tablenotemark{b}} &
\colhead{LSB Candidates\tablenotemark{b}}\\
}
\startdata
Wavelet screening & 3 & 717,124 (28,685;  20,038 to 55,374) & NA \\
Object matching & 4  & 541,086 (21,643;  14,140 to 49,328)\tablenotemark{c} & 261,046 (10,442; 7,036 to 20,330)\\
S\'ersic screening & 5 & 211,104 (8,444; 5,114 to 14,215)  & 153,552 (6,142; 3,864 to 10,733)\\
Initial GALFIT screening & 6& NA & 27,209 (1088; 686 to 1822) \\
Final GALFIT screening & 7& NA & 6044 (241; 157 to 394)\tablenotemark{d}  \\
Visual Examination &8& NA & 62 (2.48; 1 to 9)\\
\enddata
\tablenotetext{a}{Entries may include detections or candidates that are common to the two runs.}
\tablenotetext{b}{Values in parentheses denote the mean number of detections per field and the range in that value among the} target galaxies.
\tablenotetext{c}{Includes duplicates.}
\tablenotetext{d}{Duplicates deleted.}
\end{deluxetable*}

\subsection{Estimating Completeness}
\label{subsec:simulations}
\begin{figure}[h]
\includegraphics[scale=0.65]{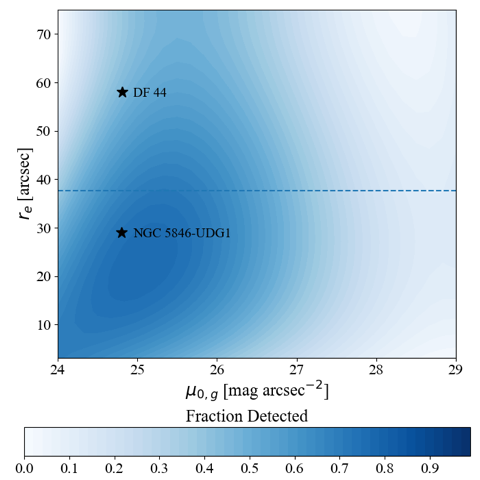}
\caption{Recovered completeness as a function of $g$-band central surface brightness and angular size on average across the survey. For guidance, we plot the locations of two well-known UDGs, DF 44 and NGC 5846-UDG1, if they are placed at the median LIGHTS distance (15.2 Mpc). The photometric data for DF 44 comes from \cite{smudges5} while that for NGC 5846-UDG1 comes from \cite{forbes19}. We also show the upper end of the $r_e$ range of the galaxies we have found with the dashed line. We are highly complete in such systems to a central surface brightness approaching $\sim 27$ mag arcsec$^{-2}$. We note, however, that the lowest surface brightness Milky Way satellite known, Antlia 2 \citep{antlia} lies at $\mu_{0,g} = 30$ mag arsec$^{-2}$, assuming $V-g = -0.3$ and $\mu_{0} = \mu_e - 1$ \citep{ji}, and is well beyond our capability to detect with this survey.}
\label{fig:completeness}
\end{figure}

We define completeness as the probability that a candidate with given structural and photometric parameters survives our pipeline.  We estimate that probability by placing simulated LSB satellites with  S\'ersic profiles and random structural and photometric properties within the image and processing them separately with the same pipeline that was used for our real sources. These are placed in a fixed grid with nodes separated by 80 arcsec in each direction, resulting in a simulation density of 2,025 deg$^{-2}$. Because we are primarily interested in the impact of $\mu_{0,g}$ and $r_e$ on completeness, we fix the other parameters ($g-r$ color, $b/a$, $n$) to their respective median values of the science candidates (0.61 mag, 0.68, and 0.81).  We sample the parameter ranges of
$23 \le\mu_{0,g}< 29$ mag arcsec$^{-2}$ and $2.7 \le r_e < 23$ arcsec. These limits are set to optimize coverage of the parameter range sampled by our candidates. 
We then extend the range of our completeness simulations to $r_e = 70$ arcsec, at lower sampling density than what is described immediately below, to broadly explore our overall detection sensitivity (Figure \ref{fig:completeness}). These values are representative across the full survey. There are fields, and areas within fields with more overlapping exposures, where we go deeper than indicated by these representative values.

To obtain enough data to adequately model completeness, we run the simulation routine 20 times which produces 339,625 simulations that fall within the image boundaries. Those sources where $>30\%$ of the pixels within one $r_e$ have NaN values are ignored.  We consider any candidate surviving the entire pipeline and lying within 4 arcsec of a simulated source to be a potential detection of that simulated source.  However, such candidates may also result from the detection of a real source in that same area. We account for such superpositions by deleting any simulation that lies within 4 arcsec of a candidate produced by the science pipeline before visual confirmation, leaving us with 338,459 simulated source of which 189,670 (56\%) survive the pipeline.  

We use these results to derive the completeness as a function of $\mu_{0,g}$ and $r_e$ using a 2\textsuperscript{nd} degree polynomial model. To derive that model we use 
the PolynomialFeatures function from the Python Scikit-learn library \citep{sklearn} and a three layer neural network implemented with Keras \citep{keras}.  For the interested reader, details on how the parameter window size and the degree of the polynomial model are selected are described in \cite{smudges2}. This procedure 
provides a good fit to the simulation results with a Coefficient of Determination, R$^2$, of 0.98. 
The resulting model (Figure \ref{fig:completeness}) is then applied to the particular parameters of each science candidate to assign each its own individual completeness estimate f$_C$.
Those values are presented in Table \ref{tab:catalog}. For three LSB candidates we either do not have the necessary data or they fall outside of the parameter space defined by the model. These are designated with f$_C = -1$ in the Table, but we adopt the mean completeness (0.56) for these three candidates when generating figures where we apply completeness corrections.

\begin{longrotatetable}
\begin{deluxetable}{llrrrrrrrrrrr}
\tablecaption{LIGHTS Low Surface Brightness Satellite Catalog}
\label{tab:catalog} 
\tablehead{ 
\colhead{SMDG Designation}&
\colhead{Parent}&
\colhead{RA}&
\colhead{Dec}&
\colhead{n}&
\colhead{r$_e$}&
\colhead{b/a}&
\colhead{PA}&
\colhead{$\mu_{0,g}$}&
\colhead{$\mu_{0,r}$}&
\colhead{r$_{e,phys}$}&
\colhead{R$_{proj}$}&
\colhead{f$_C$}\\ 
&
&
\colhead{[$^\circ$]}&
\colhead{[$^\circ$]}&
&
\colhead{[$^{\prime\prime}$]}&
&
\colhead{[$^\circ$]}&
\colhead{[mag/$(^{\prime\prime})^{2}$]}&
\colhead{[mag/$(^{\prime\prime})^{2}$]}&
\colhead{[kpc]}&
\colhead{[kpc]} 
}
\startdata
SMDG0239526-081243 &    NGC1042 &    39.96930 & $-$8.21195 &  0.85 &  8.70 &  0.9 & $-$85.1 &   27.1 &   26.4 &  0.57 &   60.4 & 0.35\\ 
SMDG0240045-082646 &    NGC1042 &    40.01880 & $-$8.44614 &  0.59 & 16.70 &  0.7 & $-$15.2 &   26.8 &   26.1 &  1.09 &   19.1 & 0.46\\ 
SMDG0240069-081343 &    NGC1042 &    40.02886 & $-$8.22864 &  0.94 & 12.53 &  1.0 &  66.0 &   24.1 &   23.3 &  0.82 &   51.0 & 0.72\\ 
SMDG0240287-081434 &    NGC1042 &    40.11961 & $-$8.24291 &  0.88 & 11.42 &  1.0 & $-$44.6 &   26.5 &   25.8 &  0.75 &   45.2 & 0.49\\ 
SMDG0240406-082308 &    NGC1042 &    40.16902 & $-$8.38542 &  0.36\tablenotemark{a} &  4.49 &  0.6 &  88.3 &   27.3 &   26.7 &  0.29 &   19.7 & 0.16\\ 
SMDG0240557-082639 &    NGC1042 &    40.23225 & $-$8.44416 &  1.09 &  5.87 &  0.8 & $-$42.6 &   27.9 &   27.4 &  0.38 &   31.0 & 0.13\\ 
SMDG0241094-081749 &    NGC1042 &    40.28910 & $-$8.29681 &  0.62 &  8.64 &  0.7 &  23.9 &   25.4 &   24.5 &  0.57 &   54.6 & 0.69\\ 
SMDG0858542+443744 &    NGC2712 &   134.72591 &    44.62899 &  0.84 &  8.83 &  0.8 & $-$5.1 &   25.0 &   24.5 &  1.29 &  160.5 & 0.75\\ 
SMDG0900028+445524 &    NGC2712 &   135.01159 &    44.92331 &  1.63 &  4.70 &  0.8 & $-$27.8 &   25.7 &   25.2 &  0.69 &   50.5 & 0.49\\ 
SMDG1046186+115918 &    NGC3368 &   161.57760 &    11.98846 &  1.34 &  3.36 &  0.6 & $-$84.1 &   24.1 &   23.5 &  0.18 &   39.4 & 0.70\\ 
SMDG1046302+114522 &    NGC3368 &   161.62565 &    11.75619 &  0.51 &  9.30 &  0.9 &  18.2 &   25.1 &   24.5 &  0.51 &   17.6 & 0.75\\ 
SMDG1047059+115243 &    NGC3368 &   161.77454 &    11.87855 &  0.52 &  7.65 &  0.7 & $-$82.0 &   26.8 &   26.2 &  0.42 &   19.7 & 0.38\\ 
SMDG1047405+120258 &    NGC3368 &   161.91868 &    12.04932 &  0.97 & 10.62 &  1.0 & $-$84.3 &   25.8 &   25.0 &  0.58 &   62.6 & 0.62\\ 
SMDG1101447+285507 &    NGC3486 &   165.43630 &    28.91870 &  0.70 &  5.57 &  0.9 &  51.5 &   24.7 &   24.1 &  0.37 &   71.2 & 0.74\\ 
SMDG1115343+143132 &    NGC3596 &   168.89298 &    14.52549 &  1.21 &  5.58 &  0.7 &  49.4 &   24.1 &   25.0 &  0.31 &   56.2 & 0.67\\ 
SMDG1126214+434408 &    NGC3675 &   171.58937 &    43.73553 &  1.06 &  9.52 &  0.5 &  87.1 &   24.0 &   23.3 &  0.78 &   45.3 & 0.66\\ 
SMDG1126532+432749 &    NGC3675 &   171.72158 &    43.46373 &  0.76 &  4.90 &  0.8 & $-$28.4 &   26.3 &   25.7 &  0.40 &   53.3 & 0.39\\ 
SMDG1132312+470704 &    NGC3726 &   173.12985 &    47.11774 &  0.98 &  6.67 &  0.8 & $-$56.5 &   27.0 &    ... &  0.44 &   39.7 & 0.32\\ 
SMDG1152340+370702 &    NGC3941 &   178.14169 &    37.11711 &  0.61 &  9.32 &  0.7 & $-$35.1 &   25.7 &   24.9 &  0.62 &   35.8 & 0.64\\ 
SMDG1155345+551658 &    NGC3972 &   178.89382 &    55.28269 &  0.82 &  4.15 &  0.8 & $-$38.7 &   27.2 &   26.6 &  0.42 &   16.5 & 0.18\\ 
SMDG1155406+552155 &    NGC3972 &   178.91909 &    55.36520 &  0.87 & 11.96 &  0.4 & $-$47.0 &   25.8 &   25.2 &  1.21 &   16.6 & 0.63\\ 
SMDG1156093+551556 &    NGC3972 &   179.03881 &    55.26550 &  0.71 &  7.54 &  0.5 &  66.0 &   24.4 &   23.8 &  0.76 &   28.9 & 0.68\\ 
SMDG1158469+473107 &    NGC4010 &   179.69539 &    47.51862 &  0.85 &  6.20 &  0.7 & $-$23.7 &   24.0 &   23.5 &  0.54 &   80.7 & 0.64\\ 
SMDG1214528+474319 &    NGC4220 &   183.72000 &    47.72185 &  1.12 &  6.66 &  0.5 &  38.0 &   25.3 &   24.7 &  0.66 &   96.9 & 0.66\\ 
SMDG1215038+473833 &    NGC4220 &   183.76590 &    47.64263 &  1.25 & 20.31 &  0.6 & $-$87.4 &   26.8 &   26.5 &  2.00 &  108.7 & 0.45\\ 
SMDG1217358+474748 &    NGC4220 &   184.39904 &    47.79678 &  0.52 &  3.37 &  0.8 &  20.9 &   26.0 &   25.7 &  0.33 &   88.7 & 0.34\\ 
SMDG1220588+085156 &    NGC4307 &   185.24491 &     8.86568 &  1.01 & 14.30 &  0.5 &   1.3 &   24.9 &   24.1 &  1.39 &  114.4 & 0.72\\ 
SMDG1222096+153910 &    NGC4321 &   185.53988 &    15.65282 &  0.74 & 15.12 &  0.8 & $-$15.0 &   26.1 &   25.4 &  1.11 &   65.8 & 0.59\\ 
SMDG1222266+084719 &    NGC4307 &   185.61079 &     8.78870 &  0.71 & 15.57 &  0.8 & $-$75.4 &   24.2 &   24.1 &  1.51 &   93.9 & 0.76\\ 
SMDG1222421+085000 &    NGC4307 &   185.67536 &     8.83311 &  0.56 & 15.03 &  0.7 & $-$59.5 &   24.6 &   24.1 &  1.46 &   90.2 & 0.74\\ 
SMDG1222509+085408 &    NGC4307 &   185.71224 &     8.90232 &  0.83 & 11.97 &  0.5 & $-$14.8 &   24.6 &   24.1 &  1.16 &   81.6 & 0.69\\ 
SMDG1222512+085046 &    NGC4307 &   185.71333 &     8.84620 &  0.68 &  8.27 &  0.9 & $-$66.3 &   27.1 &   26.8 &  0.80 &   95.0 & 0.33\\ 
SMDG1222537+160000 &    NGC4321 &   185.72376 &    16.00001 &  1.19 &  5.07 &  0.7 & $-$49.4 &   24.6 &   24.1 &  0.37 &   47.3 & 0.75\\ 
SMDG1222554+153335 &    NGC4321 &   185.73087 &    15.55973 &  0.69 & 10.07 &  0.8 &  12.8 &   25.1 &   24.3 &  0.74 &   69.6 & 0.74\\ 
SMDG1223051+155555 &    NGC4321 &   185.77118 &    15.93186 &  0.76 & 12.63 &  0.5 &  75.2 &   24.0 &   23.3 &  0.93 &   31.2 & 0.72\\ 
SMDG1248031+134329 &    NGC4689 &   192.01308 &    13.72468 &  0.67 & 17.31 &  0.6 & $-$28.1 &   24.2 &   23.6 &  1.26 &   21.1 & 0.78\\ 
SMDG1248294+134819 &    NGC4689 &   192.12234 &    13.80532 &  0.85 &  7.56 &  0.6 & $-$34.7 &   24.5 &   23.9 &  0.55 &   47.7 & 0.69\\ 
SMDG1314090+362001 &    NGC5033 &   198.53739 &    36.33372 &  1.03 &  9.71 &  0.8 & $-$29.4 &   26.3 &   25.5 &  0.90 &   98.4 & 0.52\\ 
SMDG1314205+363412 &    NGC5033 &   198.58532 &    36.56995 &  0.74 & 15.58 &  0.6 &  18.9 &    ... &   23.2 &  1.44 &   59.7 & $-1$.00\\ 
SMDG1337327+090817 &    NGC5248 &   204.38635 &     9.13815 &  0.78 &  8.01 &  0.9 &   1.6 &   24.8 &   24.1 &  0.58 &   65.7 & 0.73\\ 
SMDG1337528+085709 &    NGC5248 &   204.46989 &     8.95253 &  0.63 &  5.37 &  0.9 &  47.0 &   26.8 &   26.1 &  0.39 &   28.2 & 0.31\\ 
SMDG1504281+554100 &    NGC5866 &   226.11690 &    55.68306 &  0.74 &  3.64 &  0.4 & $-$5.0 &   25.9 &   25.3 &  0.25 &   72.9 & 0.38\\ 
SMDG1504500+553844 &    NGC5866 &   226.20822 &    55.64542 &  1.09 & 11.74 &  0.9 & $-$84.7 &   24.1 &    ... &  0.80 &   64.4 & 0.70\\ 
SMDG1505300+555200 &    NGC5866 &   226.37487 &    55.86672 &  0.79 & 37.73 &  0.5 &  63.7 &   25.6 &    ... &  2.58 &   42.7 & $-$1.00\\ 
SMDG1505523+553200 &    NGC5866 &   226.46812 &    55.53341 &  0.92 &  5.89 &  0.6 &  66.2 &   26.1 &   25.4 &  0.40 &   60.5 & 0.48\\ 
SMDG1507165+552829 &    NGC5866 &   226.81866 &    55.47484 &  0.67 &  6.24 &  0.5 & $-$46.0 &   24.6 &   24.1 &  0.43 &   76.0 & 0.74\\ 
SMDG1508054+555216 &    NGC5866 &   227.02246 &    55.87100 &  0.35\tablenotemark{a} &  4.52 &  1.0 &  89.7 &   27.9 &   27.4 &  0.31 &   61.3 & 0.10\\ 
SMDG1514241+562934 &    NGC5907 &   228.60050 &    56.49267 &  1.01 &  3.77 &  0.7 &  30.8 &   25.2 &   24.7 &  0.30 &   76.0 & 0.55\\ 
SMDG1514553+561124 &    NGC5907 &   228.73041 &    56.19002 &  0.90 &  4.31 &  0.5 & $-$65.5 &   ... &   23.1 &  0.34 &   55.8 & $-$1.00\\ 
SMDG1515425+564216 &    NGC5907 &   228.92694 &    56.70454 &  0.30\tablenotemark{a} &  3.59 &  0.5 & $-$16.3 &   27.6 &   25.9 &  0.29 &  108.5 & 0.12\\ 
SMDG1515436+561925 &    NGC5907 &   228.93171 &    56.32362 &  0.38\tablenotemark{a} &  6.26 &  0.5 & $-$31.4 &   24.2 &   23.4 &  0.50 &    6.9 & 0.68\\ 
SMDG1516240+562637 &    NGC5907 &   229.09993 &    56.44363 &  0.81 &  5.85 &  0.9 & $-$10.7 &   24.6 &   24.0 &  0.47 &   38.7 & 0.73\\ 
SMDG1516500+561315 &    NGC5907 &   229.20851 &    56.22097 &  0.66 &  3.49 &  0.6 & $-$43.0 &   25.3 &   24.8 &  0.28 &   48.7 & 0.51\\ 
SMDG1551357+621734 &    NGC6015 &   237.89891 &    62.29272 &  0.83 & 15.61 &  0.8 & $-$41.1 &   24.6 &   24.2 &  1.44 &    8.9 & 0.74\\ 
\enddata
\tablenotetext{a}{Sersic index values below 0.5 are not physically meaningful as they would imply a hole in the center of the object in 3D (Trujillo et al. 2001)}
\end{deluxetable}
\end{longrotatetable}

\section{Results}
\label{sec:results}

\begin{figure}[ht]
    \centering
    \includegraphics[scale=0.9]{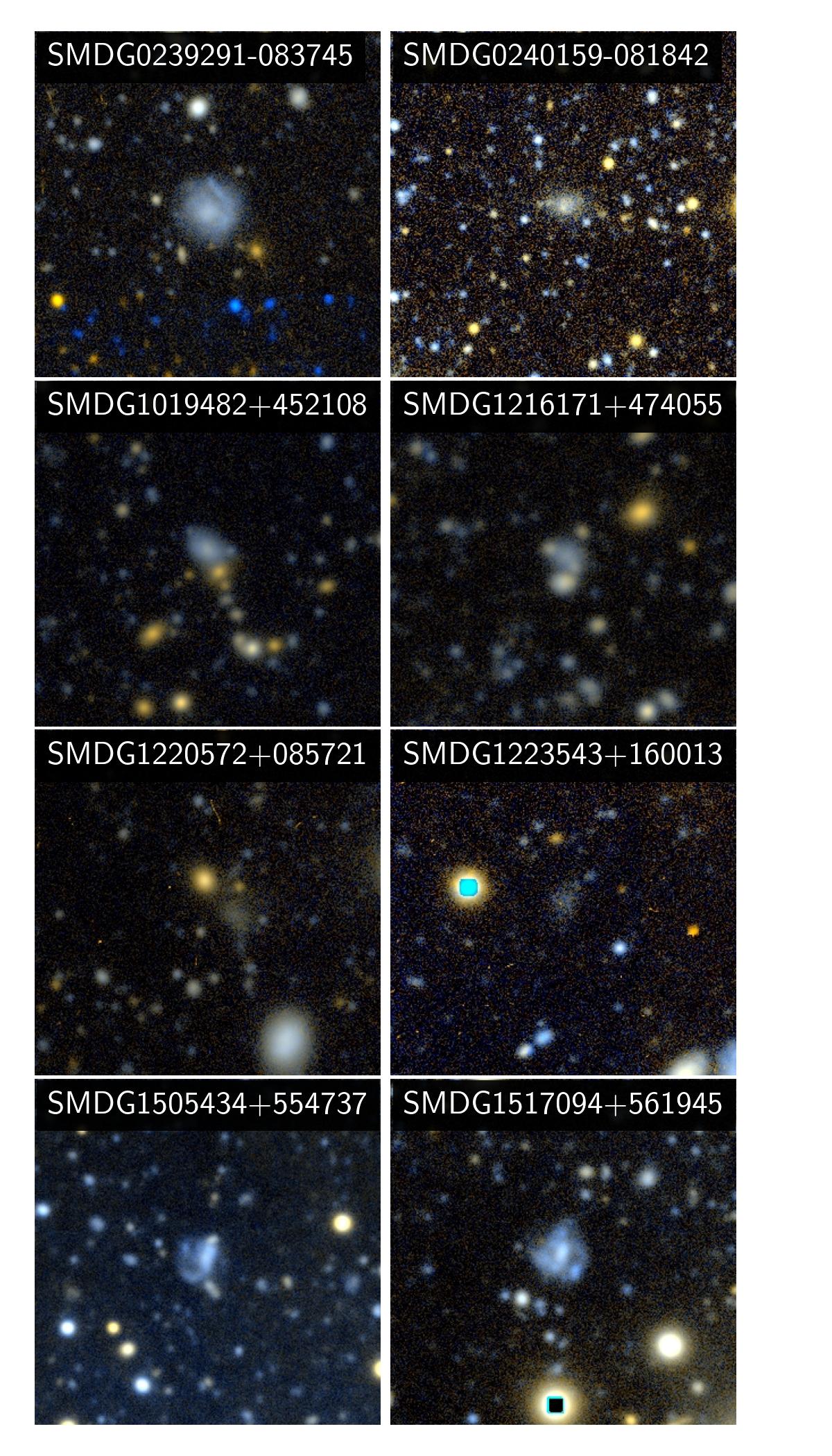}
    \caption{The eight candidates that we reject as background galaxies on the basis of visual appearance, although a few (SMDG0240159-081842, SMDG1220572+085721 and SMDG1223543+160013) might indeed be satellites. The images are $\sim$ 70 arcsec on a side and North is up.}
    \label{fig:background}
\end{figure}

\begin{figure*}[ht]
    \centering
    \includegraphics[scale=0.55]{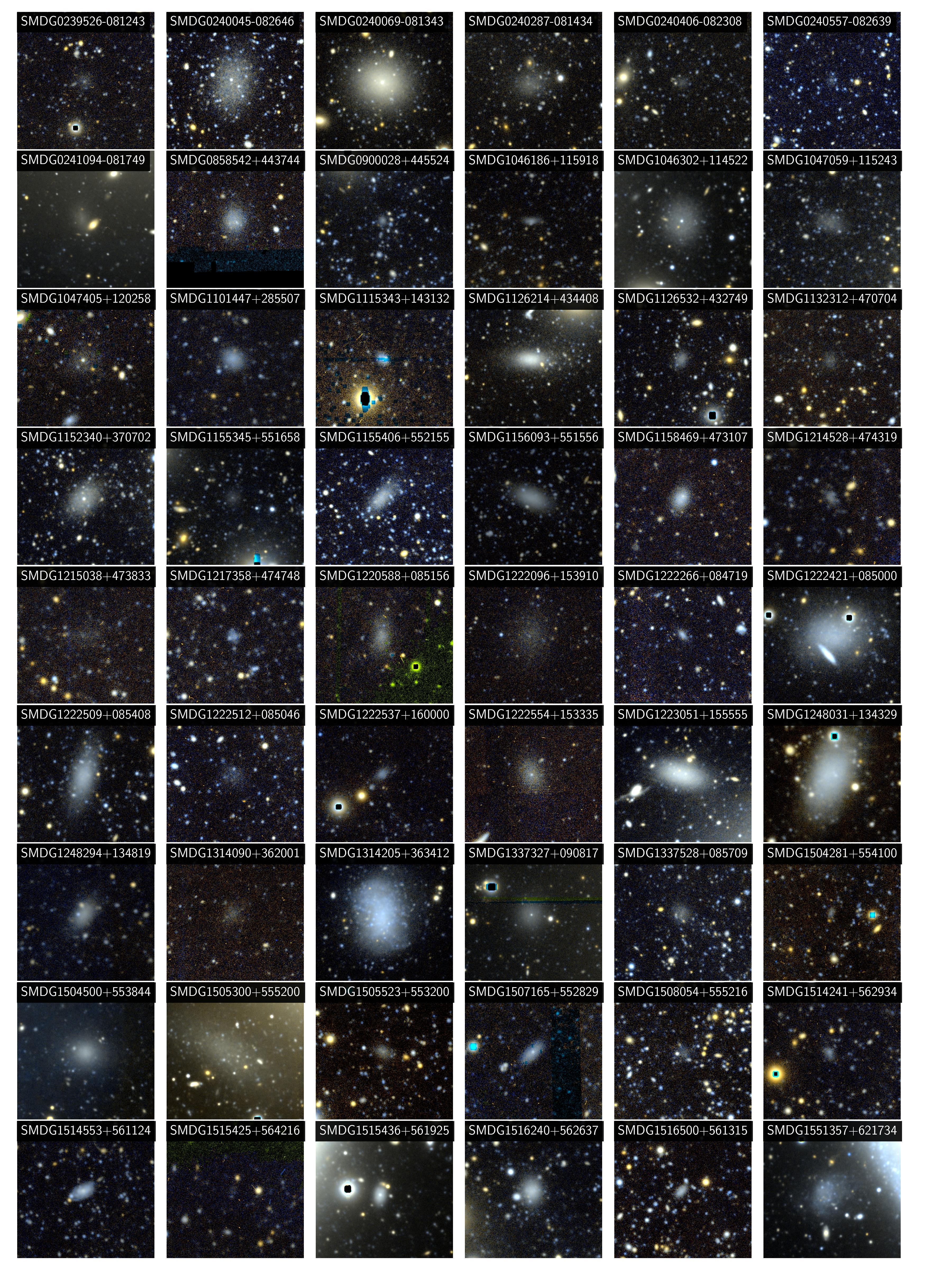}
    \caption{Low surface brightness candidate satellite galaxies identified around LIGHTS galaxies (Table \ref{tab:catalog}). Each panel is $\sim 1.5$ arcmin on a side with North at the top. The color images are created by combining $g$ and $r$ band using \texttt{astscript-color-faint-gray} \citep{2024colorimages}. }
    \label{fig:sat_mosaic}
\end{figure*}

We identify 62 low-surface brightness ($\mu_{0,g} > 24 $ mag arcsec$^{-2}$) candidate satellites around 25 target galaxies (Table \ref{tab:targets}). We reject eight of these that upon further review we consider to be background objects (Figure \ref{fig:background}). 
In some cases, these appear to have spiral structures, and are therefore likely to be background disk galaxies. In others, they are comprised of a set of small, distinct sources that are likely to be distant galaxy groups or clusters \cite[e.g.,][]{gonzalez}. These are among the smallest, highest surface brightness galaxies in our sample, which is the parameter range where one would expect to find background objects. We present the remaining 54 candidate satellites in Figure \ref{fig:sat_mosaic} and Table \ref{tab:catalog}.
Uncertainties are not included because the internally estimated errors are underestimates. Improved estimates can be obtained from simulations \citep[as done in][]{smudges5}, but uncertainties in the physical parameters, which are of primary interest here, are dominated by distance uncertainties, which are given in Table \ref{tab:targets}.

\begin{figure}[ht]
    \centering
    \includegraphics[scale=0.45]{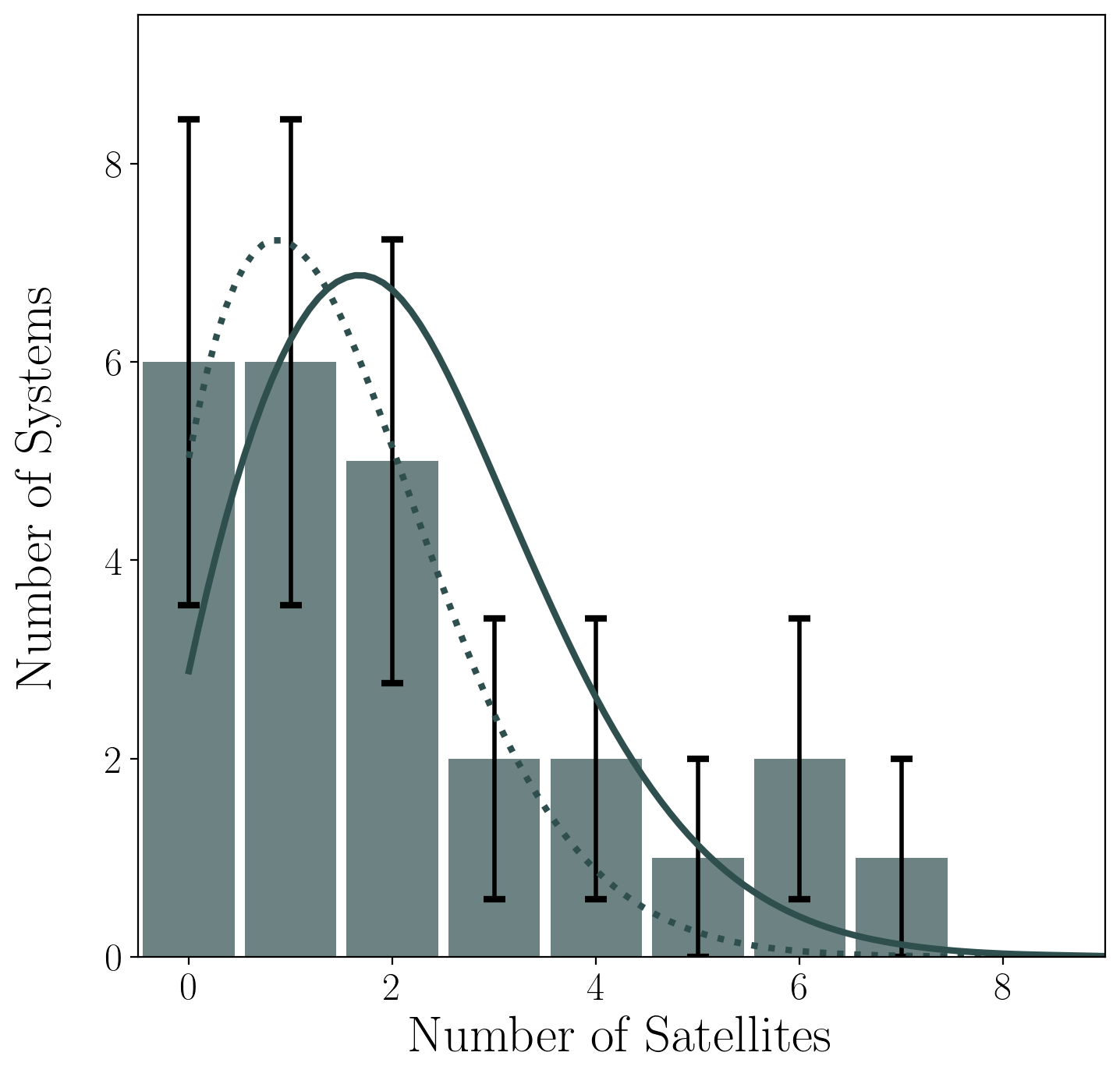}
    \caption{The distribution of target galaxies as a function of low surface brightness satellite count (not corrected for completeness). Solid line is a Poisson distribution with mean 54/25 that integrates to 25 systems. The errorbars represent $\sqrt{N}$ uncertainties only. The dotted line revisits the Poisson fit if we exclude systems with $>$ 5 satellites. The mean in this case is 1.4 LSB identified satellite candidates per host}.
    \label{fig:numbers}
\end{figure}

The number of low-surface brightness satellites candidates per parent galaxy (Table \ref{tab:targets}) varies between 0 and 7 and is reasonably reproduced by a Poisson distribution with mean 2.2 (54/25) that integrates up to 25 systems (Figure \ref{fig:numbers}). The sample size is small however and there is a hint that there are more than the expected number of systems with $\ge$ 5 satellites. Perhaps these systems are not a population of bound satellite galaxies, but rather belong either to a clustered background population or to a rich local environment if the associated parent galaxies are themselves members of a group. In two cases (NGC 1042 and  NGC 4307) the larger N$_{LSB}$ is matched with a large N$_{LEDA}$, suggesting that these are indeed found within group environments. Removing these systems from consideration, we find that the mean number of LSB satellite candidates per parent drops to 1.8 (42/23). 

A closer inspection of the numbers, however, must include consideration that our sample spans a range of distances, host galaxy stellar masses, and environments. We expect the first to affect the radial range over which we identify satellite candidates and the physical size distribution of the candidates, and the second, assuming a connection between stellar mass and total mass, to be reflected in the numbers of satellites. The third we have attempted to mitigate by not considering targets in dense environments. 

The FOV of the nearest target with the smallest FOV (NGC 3351, D = 10 Mpc, FOV = 42.7 arcmin on a side), results in a maximum radius to which we are potentially complete in our satellite search of 62 kpc. Therefore, to compare the numbers of candidate satellites among galaxies fairly, we can either limit all fields to this radius or correct for the incomplete sampling for some targets at larger radii. Considering only the candidates found at R$_{proj} < 60$  kpc, we do find a correlation between the host galaxy magnitude and the number of satellite candidates in the expected sense, but the statistical significance is at roughly only the 1$\sigma$ level. The average number of satellite candidates within 60 kpc per host in our catalog is 0.84. 

\begin{figure}[ht]
    \centering
    \includegraphics[scale=0.55]{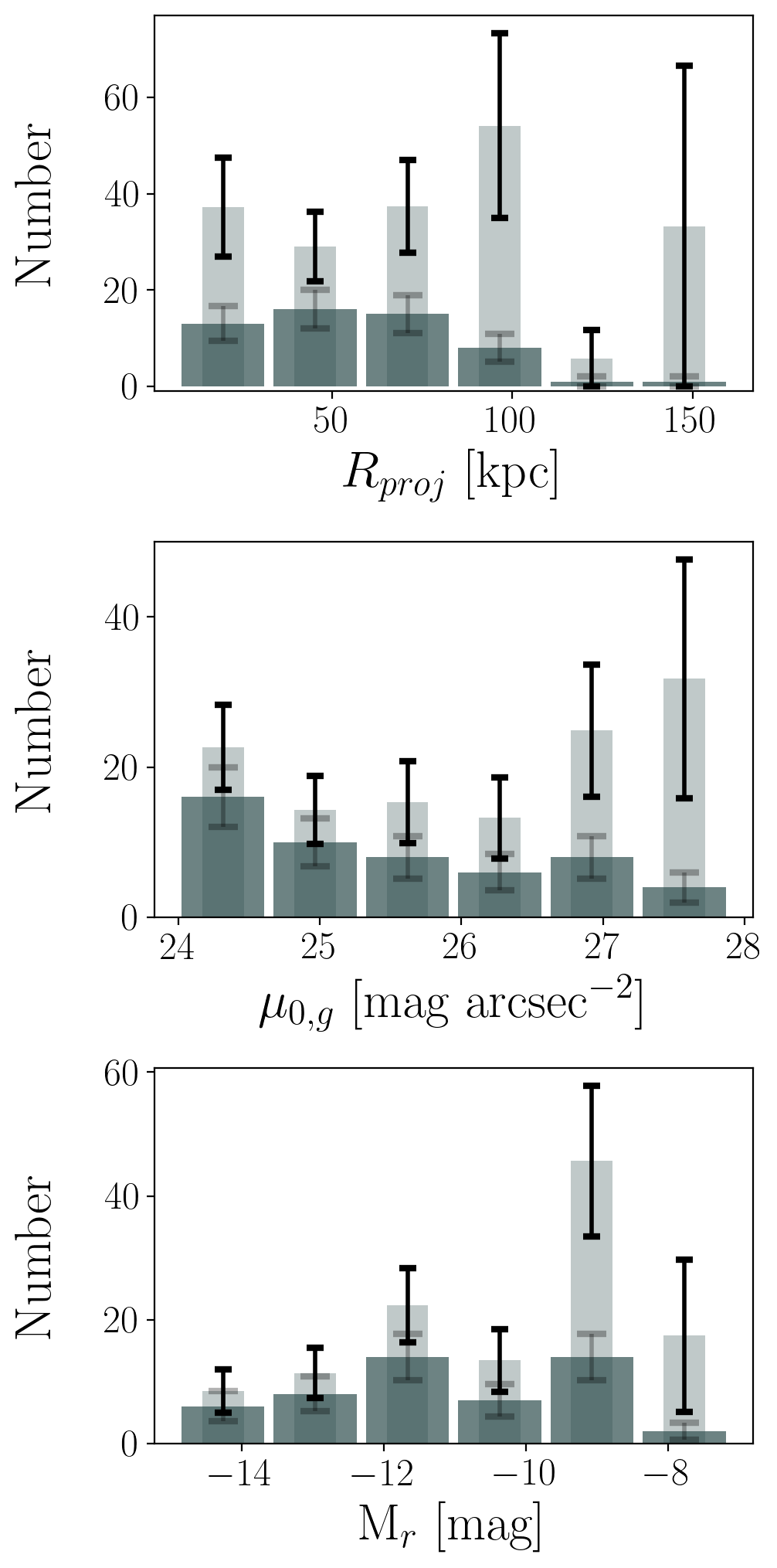}
    \caption{Basic properties of the satellite sample. Histograms represent objects in our catalog. Light/narrower bars represent completeness corrected values.}
    \label{fig:properties}
\end{figure}

Alternatively, we can probe out to a larger R$_{proj}$ and correct for those fields that do not reach these radii. In this approach, we select a radius that many of our fields reach, 100 kpc, and then simply weight the results at the larger radii by the inverse of the fraction of images that sample to that radius\footnote{This correction does not address the loss of physically smaller, lower luminosity, satellites of hosts at larger distance due to our angular size cut. Correcting for this effect requires knowledge of the satellite size and luminosity functions that we do not have.}. We refer to this as the radial completeness correction to differentiate from the photometric completeness discussed in \S3.1. When we apply both the radial and photometric corrections, we find that each host galaxy has 4 satellite candidates out to 100 kpc. Without applying the radial correction we find that each host has 3.6 candidates to this projected radius. The modest change in numbers after we apply the radial completeness correction reflects that situation that most of our fields do sample out to R$_{proj} = 100$ kpc.

Certain properties of the satellites themselves are provided in Figure \ref{fig:properties}. In the upper panel we present the radial distribution of satellites after applying both the photometric and radial completeness corrections. 
For intuition, note that if the 3-D density of satellites was proportional to $r^{-2}$ (as it would be for isothermal sphere distribution) then the number of satellites per radial bin would remain constant. We find no signs of a significant trend with radius (at R$_{proj}>100$ kpc the results are highly uncertain because only a few of our images sample that radial range).

In the middle panel, we present the central surface brightness distribution, which slightly favors brighter central surface brightnesses among the detected candidates, but extends to $\mu_{0,g} \sim 28$ mag arcsec$^{-2}$. With the photometric completeness correction applied, we find a nearly flat distribution in central surface brightness, with, at most, an allowed increase in numbers of roughly a factor of two from our bright central surface brightness limit to our faint one. We conclude that there is no vast reservoir of such undiscovered galaxies down to at least $\mu_{0,g} \sim 28$ mag arcsec$^{-2}$. 

\begin{figure*}[ht]
    \centering
    \includegraphics[scale=0.5]{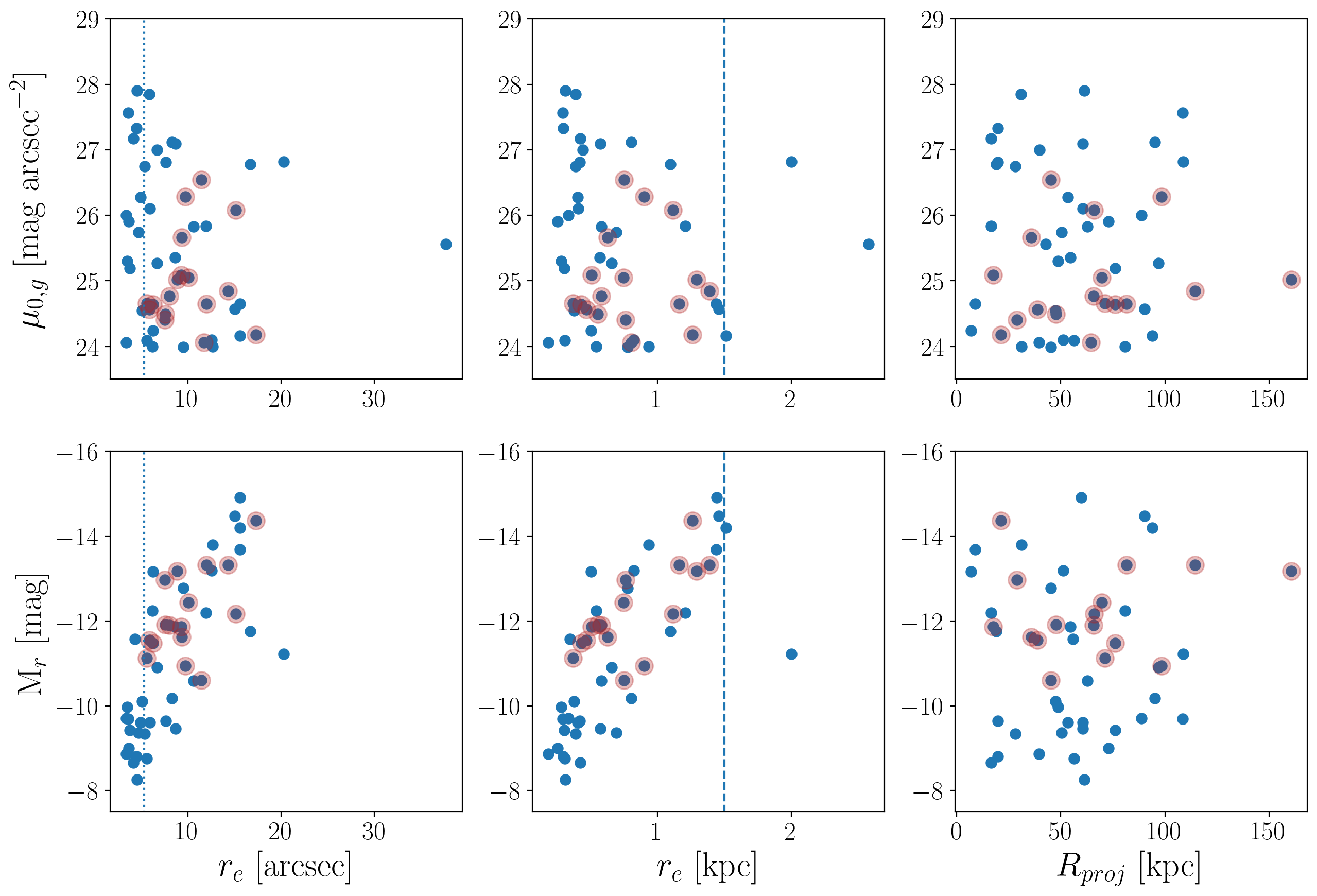}
    \caption{Distribution in satellite properties. Those galaxies also in the SMUDGes catalog \citep{smudges5} are highlighted with larger, light red circles. Vertical lines in the first column of panels mark the SMUDGes minimum angular size criterion. The vertical lines in the middle panels mark the minimum physical size that defines the UDG category. }
    \label{fig:smdg_comp}
\end{figure*}

Finally, the third panel in Figure \ref{fig:properties} shows that the distribution in luminosity for these low surface brightness candidate satellites is nearly constant between $-13 < $M$_r/$mag$ < -9$ for the detected sources. Again, the situation changes slightly after applying the photometric completeness corrections, suggesting a moderate rise in the numbers of smaller, fainter LSB satellites across the parameter range explored here. 

In summary, after applying photometric and radial completeness corrections and removing the two systems that we suspect of being contaminated with group members, we find an average of 4 LSB satellites ($24 < \mu_{0,g}/{\rm mag\ arcsec}^{-2} < 28$ and $3.2 < r_e/{\rm arcsec} < 23$) per target galaxy
out to a projected radius of 100 kpc and conclude that the number of satellites at the lower luminosities and surface brightness is likely to be larger still, but only by a moderate amount.

\section{Comparison With Previous Studies}
\label{sec:comparison}
\subsection{SMUDGes}

In Figure \ref{fig:smdg_comp} we present the comparison of our sample cross matched to the SMUDGes catalog \citep{smudges5}. Focusing on the leftmost two panels among the top set of panels in the Figure, which are directly tied to observables used in the selection, we find that there are two clear regimes where our current sample includes galaxies not in the SMUDGes set. These include those with an $r_e$ in angular units that is smaller than the limit adopted by SMUDGes (5.3 arcsec) and those with $\mu_{0,g} > 26$ mag arcsec$^{-2}$. While the former simply reflects the selection criteria of SMUDGes, the latter reflects the shallower effective surface brightness limit of the Legacy Survey data. Completeness simulations carried out for SMUDGes \citep[presented in Figure 9 of ][]{smudges2} clearly show the completeness decreasing quickly at $\mu_{0,g} \sim 26$ mag arcsec$^{-2}$. As such, data as deep as those in LIGHTS will complement the Legacy Survey data by revealing compact low surface brightness galaxies and galaxies that have central surface brightness up to  $\sim$ 2 mag fainter than the faintest in SMUDGes.

Within the regime where we expect SMUDGes to find galaxies ($r_e\ge 5.3$ arcsec and $24 < \mu_{0,g}$/mag arcsec$^{-2} < 26$) we find that LIGHTS finds 80\% more galaxies (25 vs. 14). This is in broad quantitative agreement with the relative overall completeness levels of the LIGHTS survey over this parameter range (69\%) and that for SMUDGes \citep[48\%;][]{smudges5}. 

The rightmost panels in Figure \ref{fig:smdg_comp} show how the catalog is mostly limited to R$_{proj} <$ 100 kpc, as already discussed.

\subsection{ELVES and KMTNet}

\begin{figure}[ht]
    \centering
    \includegraphics[scale=0.55]{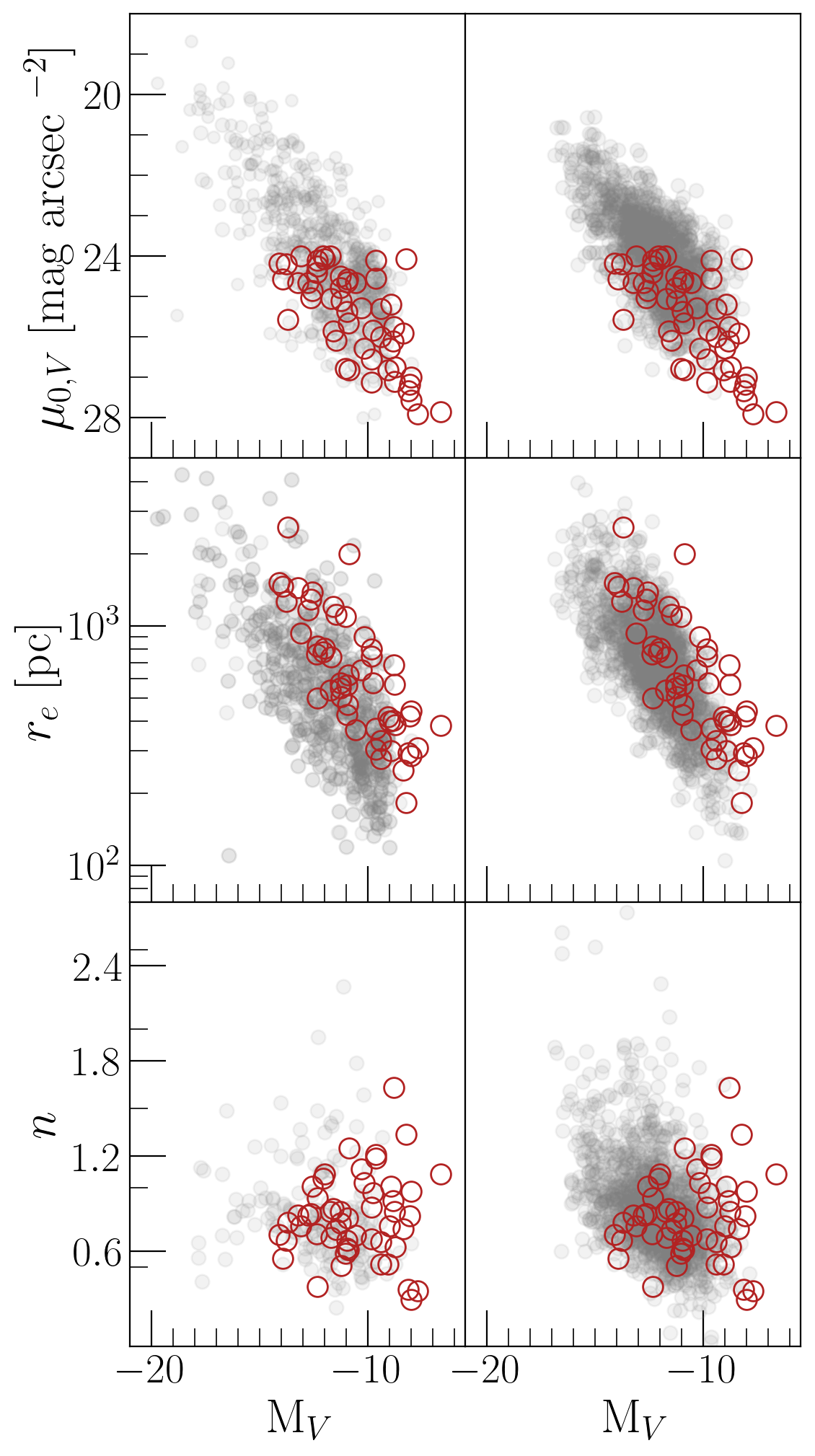}
    \caption{Comparison of properties of LIGHTS LSB satellites, open red circles, and a combination of data from the literature in filled gray circles. In the left panels the literature sample consists of data from ELVES \citep{Carlsten_2022} and KMTNet \citep{park17,park19}, while in the right panels the literature data come from MATLAS \citep{poulain}.}
    \label{fig:kmtnet_comp}
\end{figure}

Two recent surveys targeting faint satellites of nearby galaxies in intermediate to low density environments are those of \cite{Carlsten_2022} (the ELVES survey), based on CFHT imaging complemented with Legacy Survey data, and \cite{park17,park19}, based on imaging results from the KMTNet SNe survey. \cite{fan} presented a comparison of those two sets of data and we reprise that comparison including our LIGHTS detections in Figure \ref{fig:kmtnet_comp}. To make this comparison, we assume $g-V = 0$, which is likely to be off by less than 0.2 mag for blue galaxies \citep{fukugita}, but could be off by $\sim$ 0.5 mag for redder systems. Even so, neither of these differences is large enough to invalidate the qualitative comparison 
discussed here.

Our sample is consistent with those two samples once our imposed central surface brightness cut is accounted for. Our catalog extends to lower luminosity systems, by one or two mag, in at least some of our fields. Even when the depth is comparable, ELVES is limited to galaxies with distances $<$ 12 Mpc, while the LIGHTS sample only has 4 out of 25 within 12 Mpc. A nearby sample has some advantages, such as the possibility that distances can be measured using surface brightness fluctuations, but a more distant grasp provides a much larger potential sample as large area, deep imaging becomes more available.

We share one parent galaxy in common with the ELVES sample, NGC 2903. We detect 0 LSB dwarfs in our images, while \cite{Carlsten_2022} list seven satellites. Of those, however, all but two have central surface brightness significantly larger than our cut at 24 mag arcsec$^{-2}$ and the remaining two are outside of our field of view. We therefore do not find evidence of a discrepancy despite the initial impression.

\subsection{SDSS}

We do not anticipate much overlap with SDSS given our surface brightness criterion, but the potential for serendipitous spectroscopic information is of interest.
Three of our galaxies (SMDG0240069-081343, SMDG0240287-081434, and SMDG1101447+285507) match three 
SDSS galaxies (SDSS J024007.01-081344.3, SDSS J024028.61-081436.7, and SDSS J110144.69+285508.2 respectively) to within 2 arcsec. The first and last of these have associated redshift measurements in the SIMBAD database (0.9332 and 0.80471, respectively). Unfortunately, these redshift values are both implausible for these two objects which have large angular extent and low surface brightness. Of these, the first is likely an erroneous SDSS measurement (it comes with a warning in the SDSS database and its spectrum has S/N = 3.9) and the second reflects the redshift of a more distant object, a galaxy cluster, seen in projection \citep{szabo}. 

\subsection{MATLAS}

The MATLAS survey perhaps provides the most comparable data for comparison, although they obtain deep imaging primarily for elliptical galaxies \citep{matlas,matlas1}. Nevertheless, there is some overlap in target samples and \cite{poulain} provides photometric properties for the dwarf galaxies identified using MATLAS.
A general comparison between the MATLAS sample and ours is presented in Figure \ref{fig:kmtnet_comp}.

We present in Table \ref{tab:matlas} the association for ten of our galaxies with objects in the MATLAS survey. These are satellites around either NGC 3368 (designated as satellites of NGC 3379 in MATLAS), NGC 3941, NGC 3972 (designated as satellites of NGC 3998 in MATLAS), or NGC 5866. The MATLAS catalog contains numerous more sources in the vicinity of the common targets, but after eliminating galaxies at projected radii $>$ 100 kpc, where we are mostly incomplete, and galaxies with $\mu_{0,g} < 24$ mag arcsec$^{-2}$, we find only an additional 5 sources that are not in our catalog (MATLAS-734,
MATLAS-756,
MATLAS-1122,
MATLAS-2053, and
MATLAS-2083). MATLAS-734 was rejected because our visual inspection resulted in one reviewer rejecting it. Recall that in the case of a disagreement we opted to reject the detection. MATLAS-756 is beyond the edge of our image. MATLAS-1122 was rejected because our fitting latched onto a nearby star and it failed our structural parameter selection criteria. MATLAS-2053 failed the S\'ersic $n$ criterion. MATLAS-2083 failed to produce an acceptable model at all. Errors as seen in the last three objects, due presumably to nearby contaminating objects, will be captured by our completeness simulations. These three objects all appear visually to be valid satellites candidates and therefore we conclude that we found 10 of 13 satellite candidates that match our search criteria (77\%), which is a fraction entirely consistent with our estimated completeness. Note that in only one case was our visual classification the likely cause of the disagreement.

On the other side of the comparison, we identify four satellite candidates that are not included in the MATLAS catalog. These are SMDG1046186+115918, which is a satellite of NGC 3368, SMDG1155345+551658, which is a satellite of NGC 3972, and SMDG1504281+554100 and SMDG1508054+555216, which are satellites of NGC 5866. In two cases $\mu_{0,g}$ is greater than 27 mag arcsec$^{-2}$ so these are challenging sources to identify and classify properly. The other two are in a parameter range where similar systems were successfully recovered in MATLAS, so the reason for their omission is less evident, although it could be an innocuous one such as that the galaxy lies outside their survey footprint.

Our conclusion is that LIGHTS provides a comparable dwarf galaxy sample at the lowest surface brightness to that provided by MATLAS but, generally, for later type parent galaxies. LIGHTS is therefore an excellent complement to MATLAS for those interested in comparing satellites across parents of differing morphologies.

\begin{deluxetable}{rr}
\tablecaption{LIGHTS LSB Satellites Matched to MATLAS Galaxies}
\label{tab:matlas} 
\tablehead{ 
\colhead{SMDG Designation}&
\colhead{MATLAS Designation}\\
}
\startdata
SMDG1046302+114522&MATLAS-740\\
SMDG1047059+115243&MATLAS-749\\
SMDG1047405+120258&MATLAS-753\\
SMDG1152340+370702&MATLAS-1111\\
SMDG1155406+552155&MATLAS-1120\\
SMDG1156093+551556&MATLAS-1124\\
SMDG1504500+553844&MATLAS-2006\\
SMDG1505300+555200&MATLAS-2026\\
SMDG1505523+553200&MATLAS-2043\\
SMDG1507165+552829&MATLAS-2084\\
\enddata
\end{deluxetable}


\section{Discussion}
\label{sec:discussion}

\subsection{The Numbers of UDG Satellites vs. Halo Mass}

A number of studies aim to quantify the number of satellites, and in particular low surface brightness and low luminosity satellites per Milky Way analog \cite[MWA;][]{zsfw,geha,bennet19,carlsten,mao,li,goto}. The most similar, in terms of technique and selection criteria, are those of \cite{li} and \cite{goto} who found that MWAs average $0.44\pm0.05$ and $0.5\pm0.1$ UDGs, respectively. In the \cite{goto} study, MWAs were defined to have $-22 < $M$_g/$mag $< -20$ (corresponding to $-22.5 < $M$_r/$mag$ < -20.5$ for the color of the Milky Way), the satellites were in the range $-17 < $M$_g$/mag $< -13$, and the range of projected radii considered was $20 < R_{proj}/{\rm kpc} < 250$. Our study focuses on lower luminosity primaries, those that in the median are $\sim$ 1 mag fainter, and extends to satellites with lower absolute magnitudes, which will not affect the UDG counts because the UDGs are relatively bright, but does not reach out as far in projected radius.

We find three, possibly four UDGs if the one galaxies with $r_e = 1.46$ kpc (SMDG1222421+085000) is included. Although none of these were included in the SMUDGes catalog, one of these (SMDG1215038+473833) has extremely low central surface brightness $> 26.5$ mag arcsec$^{-2}$ and we would not expect SMUDGes to have included it. Two of these are in the NGC 4307 field (one is SMDG1222421+085000), which we previously noted we suspect of being in a group, and another is at R$_{proj}>100$, where we are highly incomplete (this is also the one with very low central surface brightness). If we remove those in the NGC 4307 field and the one at large radius from consideration, we conclude that we find one UDG satellite candidate in the sample of 23 target galaxies (2.5 after photometric and radial completeness corrections), for a completeness corrected frequency of 
$0.11\pm0.11$ candidate UDG satellites per host galaxy out to 100 kpc. 

\begin{figure*}[ht]
    \centering
    \includegraphics[scale=0.6]{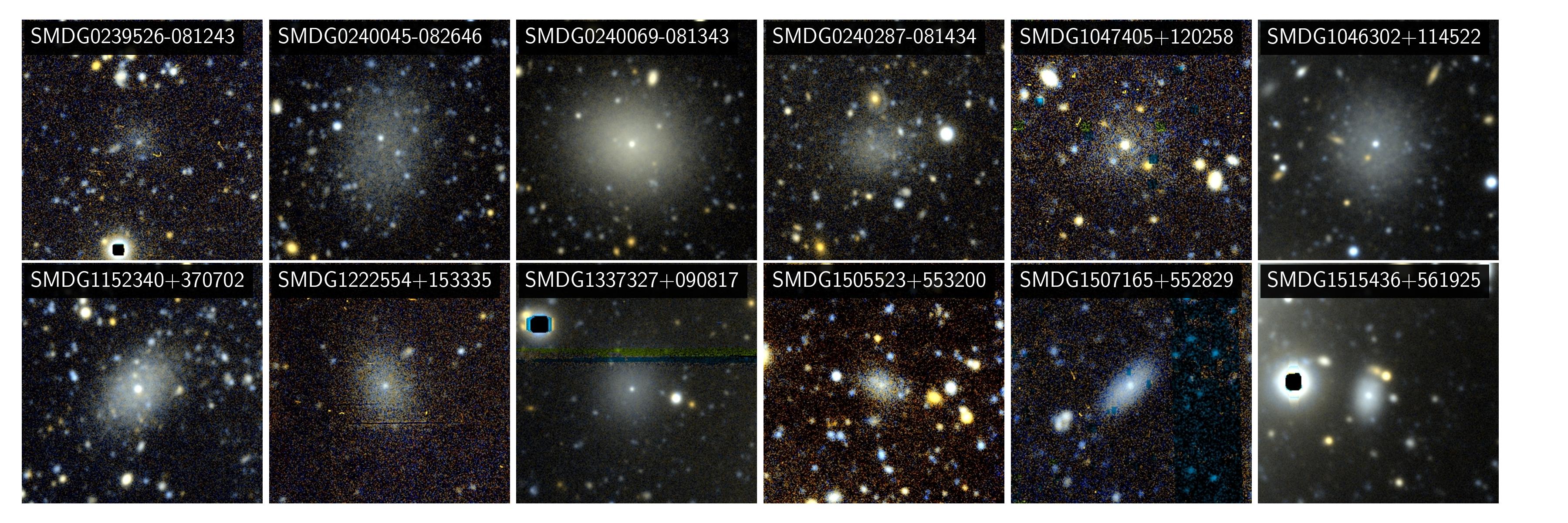}
    \caption{The twelve satellites galaxies identified visually to have nuclear star clusters. Square black regions indicate masked regions in the LIGHTS final images because they contain corrupted data. The images are $\sim$ 70 arcsec on a side and North is up.}
    \label{fig:ncs_mosaic}
\end{figure*}

We now compare our result to that of \cite{goto} for the relationship between UDGs and halo mass. Our primaries have a median magnitude of $-20.6$, which is $\sim$ one mag fainter than the midpoint of the \cite{goto} MWAs and hence a factor of 2.5 lower in stellar mass. 
At these stellar masses (log(M$_*$/M$_\odot) \sim 10.3$), the galaxies lie near the knee of the stellar mass-halo mass relation, where M$_*$/M$_h$  is roughly constant \citep{behroozi} and we adopt the approximation that the total mass-to-light ratio is comparable for our sample to that of the MWAs, resulting in the estimate that our sample has a halo mass that is 2.5$\times$ lower than that of the \cite{goto} sample. For the relationship between halo mass and the number of UDGs presented in \cite{goto} Figure 4 \citep[from ][]{karunakaran2}, we would then expect our galaxies to have 0.24 times as many UDG satellites within their virial radius as do MWAs.

Before we make the comparison to our measurement, we must account for the fact that we do not survey out to the virial radius.
Because  $r_{200} \propto M_{200}^{1/3}$, we expect the median virial radius of our galaxies to be $\sim$ 180 kpc instead of the 250 kpc of the \cite{goto} sample. If the satellite galaxy number density profile is $\propto r^{-2}$ (as discussed is consistent with the data in \S4), then we expect a factor of 100/180 UDGs when we confine ourselves to within 100 kpc, so the final expected number, extending the \cite{goto} relationship, is $0.07\pm0.01$.
This expectation is in agreement with our completion corrected measurement of $0.11\pm0.11$ UDG satellites per host and so we find no evidence for a deviation from the N$_{UDG}-$halo mass relation even when extending the relation to galaxies that are $\sim$ 1 mag fainter than MWAs\footnote{This comparison, as any between satellite samples, needs to account for the fact that these are all carried out selecting in terms of {\sl projected} radius (as stressed by \cite{goto}) and therefore includes galaxies at larger radii. Given the small number of galaxies being discussed here (i.e. one UDG), the additional complexity is not yet warranted, but should be noted.}.

\subsection{Nucleated Satellites}
\label{sec:nucleated}

We visually identify and present in Figure \ref{fig:ncs_mosaic} twelve LSB satellite candidates that host nuclear star clusters (NSCs). In a couple of cases, the fields appear sufficiently crowded with unresolved sources that the NSC may instead be an unfortunate projection of an unassociated source (e.g. SMDG0240045-082646). In general however, well centered sources are exceedingly rare and the vast majority of such sources are NSCs \citep[see][in support of this claim at shallower imaging depth]{lambert}, so we proceed for now assuming that these are all associated. For a more quantitative estimate of the likelihood of a chance superposition in the LIGHTS data we await the full source catalogs, from which we can measure the projected density of sources and assess the likelihood of a random source lying within any specific projected radius of the satellite galaxy center.

In the simplest comparison, not accounting for NSC incompleteness or relative depth and resolution differences among the samples, we compare our measurement of the NSC incidence (12/54 or 22\%) with that of the MATLAS dwarfs \citep[508/2210 or 23\%;][]{poulain}. Despite the difficulties in comparing across samples, the excellent agreement suggests that there are no gross differences either in the detection efficiencies nor in the physical properties of the galaxies themselves. The latter suggests that local evironment, in terms only as characterized by being a satellite of an early or late type galaxy, may not play a strong role in defining the NSC occupation fraction.

\section{Summary}
\label{sec:summary}

We present the current status of the LIGHTS (LBT Imaging of Galactic Halos and Tidal Structures) survey, an overview of the data processing and quality, and a catalog of low surface brightness ($\mu_{0,g} > 24$ mag arcsec$^{-2}$) satellite galaxy candidates. 

Regarding the status, we now have deep $g$- and $r$-band imaging of 25 galaxies. The target galaxies are mostly morphologically late type and in the median about a magnitude fainter than our Milky Way galaxy. They span a range of environments, although mostly low density ones, and were selected to be in low extinction regions of the sky and relatively uncontaminated by bright stars. The extinction and low stellar contamination are essential to reaching the lowest possible surface brightness.
The sample will grow slightly with time, but is mostly complete. The depth of the imaging will rival the full 10-year depth of the LSST, and so provides excellent guidance on what can be expected to become available across a large fraction of the observable sky.

We describe the data reduction process that enables us to consistently reach a 3$\sigma$ $r$-band surface brightness in areas equivalent to 10 arcsec boxes $\sim$ 30.5 mag arcsec$^{-2}$ across the sample. In particular, we rely on an iterative masking and background modeling used to achieve the final sensitivity and state-of-the-art object extraction software in Gnuastro.

We make slight modifications to the low surface brightness object detection procedure outlined by \cite{smudges,smudges2,smudges3,smudges5} to generate a catalog of 54 low surface brightness satellite galaxy candidates. These are defined to have $\mu_{0,g} > 24$ mag arcsec$^{-2}$ and $r_e > 3.2$ arcsec. Detections extend nearly to $\mu_{0,g} =28$ mag arcsec$^{-2}$, which is $\sim$ 2 magnitudes fainter than the same procedure reached on Legacy Survey images \citep{dey}. We use simulated sources to derive completeness estimates across the parameter range covered by our detected sources. These candidates are mostly confined to within 100 kpc projected radius from their parent galaxy due to the image size. 

We expect, after applying the completeness corrections, that satellites of even lower total luminosity ($M_r > -10$ mag) and lower central surface brightness ($\mu_{0,r} > 27.5$ mag arcsec$^{-2}$) to be more prevalent. Over the parameter range we explore, each host (excluding those that are in overdense regions, apparently groups) has nearly 4 LSB satellites to a projected radius of 100 kpc. These objects are mostly just at or beyond the reach of spectroscopic surveys of LSB galaxies \citep[e.g.,][]{kadowaki21} unless they are H I rich or have ongoing star formation. However, they are sufficiently close to us, if truly associated with their host galaxies, that other distance estimators utilizing resolved stars or the globular cluster luminosity function could be exploited with space-based observatories. For the fainter, smaller satellites that might have fewer luminous stars or globular clusters even these methods may fail.

In our catalog, there are 3, possibly 4, ultra-diffuse galaxies (UDGs; $r_e > 1.5$ kpc). This allows us to explore and possibly extend the relationship between the number of UDGs and the host halo mass found for more massive systems \citep{vanderburg,rt,mp,karunakaran2,goto}. Despite differences in sample selection, we find that our results are in agreement with the extrapolation of the mean relationship used by \cite{karunakaran2} to galaxies that are a magnitude fainter than the Milky Way. 

We visually identify 12 galaxies in our catalog that host a nuclear star cluster (NSC). The occupation fraction for the sample (12/54) is in excellent agreement with that found among satellites of early type galaxies \citep{poulain} and so suggests that the morphological type of the parent galaxy plays at most a limited role in the NSC satellite occupation fraction. 

The LIGHTS sample provides exquisite data for the exploration of the halo stellar populations, whether they be diffusely distributed, in tidal features, or, as described here, still part of a low surface brightness satellite galaxy. Comparison with state-of-the-art published studies shows that LIGHTS matches or supersedes the sensitivity of those and/or provides data for a different category of parent galaxy. It is a resource for current study, but also provides direct guidance on the potential of upcoming deep imaging surveys such as the LSST that will be carried out using the Rubin telescope. 

\begin{acknowledgments}

DZ acknowledges financial support from NSF AST-1713841 and AST-2006785. 
The authors thank the anonymous referee for a careful and thoughtful reading of the manuscript. An allocation of computer time from the UA Research Computing High Performance Computing (HPC) at the University of Arizona and the prompt assistance of the associated computer support group is also gratefully acknowledged. We acknowledge the usage of the HyperLeda database (http://leda.univ-lyon1.fr).
IT acknowledges support from the ACIISI, Consejer\'{i}a de Econom\'{i}a, Conocimiento y Empleo del Gobierno de Canarias and the European Regional Development Fund (ERDF) under grant with reference PROID2021010044 and from the State Research Agency (AEI-MCINN) of the Spanish Ministry of Science and Innovation under the grant PID2022-140869NB-I00 and IAC project P/302302, financed by the Ministry of Science and Innovation, through the State Budget and by the Canary Islands Department of Economy, Knowledge and Employment, through the Regional Budget of the Autonomous Community.
RIS acknowledges the funding by Governments of Spain and Arag\'on through FITE and Science Ministry (PGC2018-097585-B-C21, PID2021-124918NA-C43). MM acknowledges support from the Project PCI2021-122072-2B, financed by MICIN/AEI/10.13039/501100011033, and the European Union ``NextGenerationEU"/RTRP. 
JR acknowledges funding from University of La Laguna through the Margarita Salas Program from the Spanish Ministry of Universities ref.~UNI/551/2021-May 26, and under the EU Next Generation. 

\end{acknowledgments}

\software{
Astropy              \citep{astropy1, astropy2},
astroquery           \citep{astroquery},
Astrometry.net       \citep{astrometry.net},
GALFIT               \citep{peng},
Gnuastro             \citep{gnuastro,gnuastro2},
keras                \citep{keras},
lmfit                \citep{newville},
Matplotlib           \citep{matplotlib},
NumPy                \citep{numpy},
pandas               \citep{pandas},
sep                  \citep{sep},
Source Extractor     \citep{bertin},
SCAMP                \citep{scamp2006},
SciPy                \citep{scipy1, scipy2},
SWarp                \citep{Swarp}
}

\bibliography{references.bib}{}
\bibliographystyle{aasjournal}

\end{document}